\documentclass[11pt]{article}

\usepackage{verbatim}
\usepackage{amsmath}
\usepackage{amsfonts}
\usepackage{amssymb}
\usepackage{bbm}
\usepackage{latexsym}
\usepackage{lscape} 
\usepackage{epsfig}
\usepackage{color}
\usepackage{graphicx}
\usepackage{hyperref}

\usepackage{subfig}
\usepackage{slashbox}
\usepackage{multirow}
\usepackage{mathrsfs}

\usepackage{amscd}
\usepackage{cite}
\usepackage{units}
\usepackage{feynmp}

\renewcommand{\d}{\mathrm{d}}

\newcommand{\dd}[2][]{\ensuremath{\frac{\d #1}{\d #2}}}
\newcommand{\fd}[2][]{\ensuremath{\frac{\delta #1}{\delta #2}}}
\newcommand{\pp}[2][]{\ensuremath{\frac{\partial #1}{\partial #2}}}

\DeclareMathSymbol{\mg}{\mathrel}{symbols}{"1D}

\newcommand{\bnabla}{\ensuremath{\overline{\nabla}}}

\newcommand{\ga}{\alpha}
\newcommand{\gb}{\beta}
\renewcommand{\gg}{\gamma}
\newcommand{\gd}{\delta}
\renewcommand{\ge}{\epsilon}

\newcommand{\gf}{\phi}

\newcommand{\gx}{\xi}
\newcommand{\gm}{\mu}

\newcommand{\gl}{\lambda}

\newcommand{\gth}{\theta}

\newcommand{\gs}{\sigma}

\newcommand{\go}{\omega}

\newcommand{\gp}{\pi}
\newcommand{\gps}{\psi}
\newcommand{\get}{\eta}

\newcommand{\gG}{\Gamma}
\newcommand{\gD}{\Delta}
\newcommand{\gF}{\Phi}

\newcommand{\gX}{\Xi}
\newcommand{\gL}{\Lambda}

\newcommand{\gTh}{\Theta}
\newcommand{\gO}{\Omega}

\newcommand{\gPs}{\Psi}

\newcommand{\cD}{{\cal D}}
\newcommand{\cE}{{\cal E}}

\newcommand{\cG}{{\cal G}}

\newcommand{\cN}{{\cal N}}
\newcommand{\cO}{{\cal O}}

\newcommand{\cS}{{\cal S}}

\newcommand{\cU}{{\cal U}}

\newcommand{\ua}{{\underline a}}
\newcommand{\ub}{{\underline b}}
\newcommand{\uc}{{\underline c}}
\newcommand{\ud}{{\underline d}}

\newcommand{\tA}{{\widetilde A}}

\newcommand{\uga}{{\underline \alpha}}
\newcommand{\ugb}{{\underline\beta}}
\newcommand{\ugg}{{\underline\gamma}}

\newcommand{\Id}{\text{\small 1}\hspace{-3.5pt}\text{1}}

\newcommand{\Slashed}{\hspace{-1.4ex}/\hspace{.2ex}}

\newcommand{\ra}{\rightarrow}

\newcommand{\der}{\partial}

\newcommand{\dsp}{\displaystyle}

\newcommand{\half}{\frac 12 }

\newcommand{\Kh}{K\"{a}hler}
\newcommand{\beq}{\begin{equation}}
\newcommand{\eeq}{\end{equation}}
\newcommand{\barr}{\begin{array}}
\newcommand{\earr}{\end{array}}
\newcommand{\equ}[1]{\begin{gather} #1 \end{gather}}
\newcommand{\equa}[1]{\begin{align} #1 \end{align}}

\newcommand{\tabu}[2]{\begin{tabular}{#1} #2 \end{tabular}}
\newcommand{\arry}[2]{\begin{array}{#1} #2 \end{array}}

\newcommand{\non}{\nonumber}
\newcommand{\sfrac}[2]{\mbox{$\frac{#1}{#2}$}}
\newcounter{oldcounter}

\newcommand{\bder}{\bar\partial}
\newcommand{\bff}{{\bar f}}
\newcommand{\bg}{{\bar g}}
\newcommand{\bh}{{\bar h}}

\newcommand{\bk}{{\bar k}}

\newcommand{\bm}{{\bar m}}

\newcommand{\bz}{{\bar z}}

\newcommand{\bB}{{\overline B}}
\newcommand{\bC}{{\overline C}}
\newcommand{\bD}{{\overline D}}

\newcommand{\bH}{{\overline H}}

\newcommand{\bJ}{{\overline J}}
\newcommand{\bK}{{\overline K}}

\newcommand{\bM}{{\overline M}}

\newcommand{\bR}{{\overline R}}

\newcommand{\bW}{{\overline W}}
\newcommand{\bX}{{\overline X}}
\newcommand{\bY}{{\overline Y}}

\newcommand{\bgf}{{\bar\phi}}

\newcommand{\bgm}{{\bar\mu}}

\newcommand{\bgl}{{\bar\lambda}}

\newcommand{\bgth}{{\bar\theta}}

\newcommand{\bgps}{{\bar\psi}}

\newcommand{\bgG}{{\overline\Gamma}}

\newcommand{\bgF}{{\overline\Phi}}

\newcommand{\bgL}{{\overline\Lambda}}

\newcommand{\bgO}{{\overline\Omega}}

\newcommand{\bgPs}{{\overline\Psi}}

\newcommand{\tgG}{{\tilde \Gamma}}

\newcommand{\Real}{\mathbbm{R}}

\newcommand{\ba}[2]{\[\begin{array}{#2}\label{#1}}
\newcommand{\ea}{\end{array}\]}
\newcommand{\be}{\begin{equation}}
\newcommand{\ee}{\end{equation}}
\newcommand{\bea}{\begin{eqnarray}}
\newcommand{\eea}{\end{eqnarray}}

\newcommand{\derl}{{\partial_{_L}}}
\newcommand{\derr}{{\partial_{_R}}}

\begin{document}
\unitlength = 1mm
\begin{fmffile}{papermf} 

\thispagestyle{empty}

\begin{flushright}
 LMU-ASC 21/12 
\\[3ex]
\end{flushright}
\vskip .5cm
\begin{center}
{\Large {\bf 
Super Weyl invariance: BPS equations from heterotic worldsheets 
} 
}
\\[0pt]

\bigskip
\bigskip {\large
{\bf Stefan Groot Nibbelink
}\footnote{
E-mail: Groot.Nibbelink@physik.uni-muenchen.de},
{\bf Leonhard Horstmeyer
}\footnote{
E-mail: Leonhard.Horstmeyer@physik.uni-muenchen.de}
\bigskip }\\[0pt]
\vspace{0.23cm}
{\it 
Arnold Sommerfeld Center for Theoretical Physics,\\
~~Ludwig-Maximilians-Universit\"at M\"unchen, Theresienstrasse 37, 80333 M\"unchen, Germany
 \\} 
\end{center}

\subsection*{\centering Abstract}

It is well--known that the beta functions on a string worldsheet correspond to the target space equations of motion, e.g.\ the Einstein equations. 
We show that the BPS equations, i.e.\ the conditions of vanishing supersymmetry variations of the space--time fermions, can be directly derived from the worldsheet.  
To this end we consider the RNS--formulation of the heterotic string with (2,0) supersymmetry, which describes a complex torsion target space that supports a holomorphic vector bundle. 
After a detailed account of its quantization and renormalization, we establish that the cancellation of the Weyl anomaly combined with (2,0) finiteness implies the heterotic BPS conditions: 
At the one loop level the geometry is required to be conformally balanced and the gauge background has to satisfy the Hermitean Yang--Mills equations.

\newpage 
\setcounter{page}{1}

\section{Introduction}
\label{sc:intro}

The heterotic string~\cite{Gross:1984dd,Gross:1985fr} provides a fascinating arena for the study of phenomenologically interesting string vacua. To obtain $\cN=1$ supersymmetric models in four dimensions one often considers string compactifications on Calabi--Yau (CY) manifolds with holomorphic vector bundles~\cite{Candelas:1985en,gsw_2}. The gauge background has to satisfy the Hermitean Yang--Mills (HYM) equations. These equations can be solved provided that the corresponding vector bundle is stable~\cite{Donaldson:1985,Uhlenbeck:1986}. In fact, CY compactifications are guaranteed to be applicable only for the so--called standard embedding, in which the gauge connection is set equal to the spin--connection. For other gauge fluxes the compactification manifold often possesses torsion~\cite{Strominger:1986uh,Hull:1986kz}. To find solutions to this so--called Strominger system is a non--trivial task, see e.g.~\cite{Dasgupta:1999ss,Becker:2002sx,Gillard:2003jh,Fu:2006vj,Becker:2006et,Andreas:2010qh}.

The conditions to preserve $\cN=1$ supersymmetry in four dimensions are obtained by setting the supersymmetry variations of the gravitino, dilatino and gaugino to zero. These supersymmetry transformations leave the action of ten dimensional super Yang--Mills coupled to supergravity invariant~\cite{Bergshoeff:1982um,Bergshoeff:1989de,Chemissany:2007he}. There are essentially two ways discussed in the literature to compute this  effective action and its corrections from string theory. One approach is to compute appropriate string scattering amplitudes and compare them with the interactions which the effective action describes~\cite{Metsaev:1987ju,Metsaev:1987zx}. The other approach is to determine the target space equations of motion for the bosonic fields by computing the beta functions on the worldsheet~\cite{Friedan:1980jf,Friedan:1980jm,AlvarezGaume:1980dk,Callan:1985ia}. These equations of motion can be lifted to an effective action~\cite{Hull:1987yi}. In either case, the string derivation of the conditions for $\cN=1$ supersymmetry in four dimensions is rather indirect.

The main aim of this paper is to show that these conditions for unbroken supersymmetry can be directly obtained by worldsheet considerations. To this end we use the Ramond--Neveu--Schwarz (RNS) formulation to describe the propagation of the heterotic string in a bosonic target space field background~\cite{Gross:1984dd,Gross:1985fr}. The corresponding Non--Linear Sigma Model (NLSM) on the heterotic worldsheet possesses at least (1,0) supersymmetry. We assume throughout the paper that the worldsheet has (2,0) supersymmetry, so that the target space of the string is a complex manifold which generically possesses torsion~\cite{Hull:1985jv}. We study the renormalization of this NLSM using manifest (2,0) superspace techniques~\cite{Dine:1986by,Evans:1986tc,Brooks:1986uh}. (Recently some quantum aspects of (2,0) theories have been studied in~\cite{Cui:2010si,Cui:2011rz}.) This allows us to study the conditions for finiteness~\cite{Hull:1985zy}, i.e.\ vanishing beta functions, in a fully supersymmetric fashion.

However, finiteness is not the critical issue to obtain a sensible string theory; Weyl invariance is~\cite{Hull:1985rc}\footnote{Under some general assumptions scale invariance implies Weyl invariance~\cite{Polchinski:1987dy}, however, insisting on Weyl invariance does not require any further assumptions and only use information in a local patch.}: By requiring Weyl invariance at the quantum level certain ambiguities in the beta functions are removed and the resulting conditions can be interpreted as the target space equations of motions with a dilaton background included. We find that by combining the conditions that the (2,0) NLSM is finite with Weyl invariance at the quantum level directly results in the conditions for unbroken target space supersymmetry on the bosonic background fields.

\subsubsection*{Paper overview}

Section~\ref{sc:Classical} introduces (2,0) superfields to describe an heterotic NLSM with torsion and an arbitrary holomorphic vector bundle. The classical symmetries of this theory are identified, with a specific emphasis on its Weyl invariance. 
Section~\ref{sc:Quantum} defines the quantum effective action as a path integral from which the Feynman rules can be read off. To ensure that this definition is covariant we introduce an holomorphic normal coordinate expansion for the quantum fields. 
The renormalization of this theory is discussed in Section~\ref{sc:Renorm}. The conditions for (2,0) finiteness of the NLSM are derived. In particular, it is shown that the Weyl invariance combined with (2,0) finiteness at the one--loop level correspond to the conditions of vanishing fermion variations in target space. 
Section~\ref{sc:Concl} summarizes our main findings. 

Three Appendices have been included in this paper: Appendix~\ref{sc:Conventions} collects our (2,0) superspace conventions. A number of useful identities for torsion connections can be found in Appendix~\ref{sc:Connections}. Appendix~\ref{sc:DimRed} describes a specific adaptation of the dimensional regularization scheme.

\renewcommand{\contentsname}{Table of content}

\tableofcontents

\newpage

\section{Classical (2,0) non--linear sigma models}
\label{sc:Classical}

The worldsheet theory of the heterotic string with $\cN=1$ target space supersymmetry is described by a $(2,0)$ NLSM~\cite{Hull:1985jv}. A general conformal action for the NLSM is constructed out of chiral and fermi chiral superfields. By considering the component form of the action shows that it describes a complex torsion manifold with a holomorphic vector bundle. The theory is invariant under target space diffeomorphisms and gauge transformations. Finally, super Weyl transformations can be discussed when the NLSM has been coupled to (2,0) worldsheet supergravity.

\subsection{Superfields}
\label{sc:Superfields}

The description of a (2,0) NLSM requires the introduction of chiral and Fermi superfields. We employ (2,0) superspace methods~\cite{Dine:1986by,Brooks:1986uh,Evans:1986tc,Distler:1992gi,Witten:1993yc} throughout this paper; our conventions are collected in the Appendix~\ref{sc:Conventions}.

\subsubsection*{Chiral multiplets}

We introduce chiral superfields $\gf^a$ and their conjugates, the so--called anti--chiral superfields, $\bgf^\ua$ via the constraints
\equ{ 
\bD_+\gf^a = D_+\bgf^\ua = 0~,
\label{chiralcond}
}
with the super covariant derivatives $D_+$ and $\bD_+$ are defined in \eqref{superD}. The indices $a$ label the coordinates of the ten dimensional space--time in a complex basis. The components of the chiral multiplets are recovered from the expansion in the Grassmann variables
\equ{
z^a=\gf^a|~, 
\qquad 
\gps^a=\frac{1}{\sqrt{2}}D_+\gf^a|~,
\label{chiralcomponents}
}
where $|$ denotes that we have set $\gth^+=\bgth^+=0$. The scalar components $z^a$ define the complex coordinate fields of the target space; the right--moving fermions $\gps^a$ are their supersymmetric partners.

\subsubsection*{Fermi multiplets}

The Fermi superfields $\gL^\ga$ and $\bgL^\uga$ are fermionic (anti--)chiral superfields, 
\equ{ 
\bD_+\gL^\ga = D_+\bgL{}^\uga = 0~, 
\label{fermichiralcond}
}
$\ga = 1,\ldots, 16$. The components of the Fermi multiplets are defined by
\equ{
\gl^\ga=\gL^\ga|~,
\qquad 
h^\ga=\frac{1}{\sqrt{2}}D_+\gL^\ga|~.
\label{fermichiralcomponents}
}
The left--moving fermions $\gl^\ga$ generate the target space gauge group. The scalars $h^\ga$ are auxiliary fields which can be integrated out by their algebraic equations of motion.

\subsection{Classical action}
\label{sc:action} 

\begin{table} 
\begin{center}
\tabu{|c||c|c|c|c|c|c|c|}{
\hline
Quantity                & $\derl$ & $\derr$ & $\d\gs_L$ & $\d\gs_R$ & $\d^2\gth^+$ & $\gf$ &   $\gL$   \\\hline\hline 
Conf.\ weights        &  $(1,0)$ & $(0,1)$  & $(-1,0)$ & $(0,-1)$   & $(1,0)$  &  $(0,0)$    & $(0,\half)$  \\\hline
}
\end{center}
\caption{This Table indicate the conformal weights of the derivatives, the integral measures and the superfields which appear in a $(2,0)$ NLSM. 
\label{tb:Scalings}}
\end{table} 

In order to construct a general scale invariant NLSM we consider the left-- and right--moving scaling dimensions or conformal weights of the various quantities as given in Table~\ref{tb:Scalings}. A supersymmetric (2,0) action can be written as a full superspace integral $\int\d^2\gs\d^2\gth^+$. This implies that the integrand is  required to have conformal weights $(0,1)$. Consequently, the only admissible terms contain $\derr \gf$, $\derr \bgf$ or Fermi bilinears $\bgL \gL$, $\gL\gL$ and $\bgL\,\bgL$ times arbitrary functions of $\gf$ and $\bgf$. Therefore, the full action $S = S_\gf + S_\gL$ takes the general form~\cite{Dine:1986by}:
\begin{subequations} 
\equa{
S_\gf = & \dsp 
\frac i4 \int\d^2\gs\d^2\gth^+\,
\Big\{ \bK_\ua(\gf,\bgf)\, \derr\bgf^\ua - K_a(\gf,\bgf) \, \derr\gf^a \Big\}~, 
\label{ChiralAction} 
\\[1ex] 
S_\gL = & \dsp 
 -\frac 12  \int\d^2\gs\d^2\gth^+\,\Big\{
 \bgL^\uga \,N_{\uga\gb}(\gf,\bgf) \, \gL^\gb 
+\half\, \gL^\ga \,M_{\ga\gb}(\gf,\bgf) \, \gL^\gb 
+\half\, \bgL^\uga \, \bM_{\uga\,\ugb}(\gf,\bgf) \, \bgL^\ugb  
\Big\}~. 
\label{FermiAction} 
}
\end{subequations} 
A factor of $\ga '$ has been absorbed into the chiral superfields and can be reintroduced by a scaling $z \ra z/\sqrt{4 \gp \ga'}$. This defines the most general $(2,0)$ action; a superpotential is not admissible: As a  superpotential is integrated over a chiral subspace, it has to be linear in the Fermi superfields. In two dimensions this would require an operator with weights $(\half,\half)$, but no such operator exists. Hence, a superpotential can only be written down with an explicit mass parameter, but this breaks scale invariance.

If we write $\gf = (\gf^a)$ as a column vector and $\bgf = (\bgf^\ua)$ and a conjugate row vector, we can define the row and column vectors $K = (K_a)$ and $\bK = (\bK_\ua)$. In addition we write $\gL = (\gL^\ga)$ as a column vector and $\bgL = (\bgL^\uga)$ as a row vector, so that the matrix functions $N = [N_{\uga\ga}]$, $M = [M_{\ga\gb}]$ and $\bM = [\bM_{\uga\ugb}]$ define rank $(1,1)$-- $(2,0)$-- and $(0,2)$--tensors, respectively. The complete classical action $S$ can be compactly written as 
\equ{
S = \int \d^2\gs \d^2 \gth^+\, \Big\{
\frac{i}{4} \Big( \derr\bgf\, \bK - K\,\derr\gf \Big) 
- \frac{1}{2}\Big( \bgL\, N\, \gL
+\frac{1}{2}\, \gL^T \, M\,\gL
+\frac{1}{2}\, \bgL\,\overline{M} \,\bgL{}^T \Big) 
\Big\}~. 
\label{ClAction}
}
To understand the target space interpretation of this action we exploit its component form below.

\subsubsection*{Torsion complex manifold}

The scalar field part of the component action does not have a unique representation 
\equ{ 
S_\gf \supset \int\d^2\gs\, 
\Big\{ \half\, G_{\ua a}
\big(\derl z^a \derr \bz^\ua  + \derl \bz^\ua \derr z^a\big) 
+ 
 \frac{1 - \gb}2\, B_{\ua a}
\big(\derl z^a \derr \bz^\ua  - \derl \bz^\ua \derr z^a\big) + 
\non \\[2ex] 
 - \half\, \gb\, B_{a b}\, \derl z^b \derr z^a 
- \half\, \gb\, \bB_{\ua\, \ub} \, \derl \bz^\ub\derr \bz^\ua 
\Big\}~,
\label{ScalarAcChiral}
}
as it depends on the choice of a real parameter $\gb$. The derivation of this expression shows that the functions $K_a$ and $\bK_\ua$ are pre--potentials for both a metric $G$ and a two--form field $B_2$. The metric $G = [G_{\ua a}]$ is an Hermitian metric of some complex manifold, as its only non--vanishing components read  
\equ{ 
G_{\ua a} = \half \big(\bK_{\ua, a} + K_{a, \ua} \big)~. 
\label{Metric} 
}
The components of the inverse metric are denoted with upper indices: 
$[G^{-1}]^{b \ua} = G^{b\ua}$. The components of the two--form 
$B_2 = B_{\ua b}\, \d\bz^\ua\wedge \d z^b +
 \half B_{a b}\, \d z^a\wedge \d z^b+
 \half B_{\ua\, \ub}\, \d \bz^\ua \wedge \d \bz^\ub$ 
are expressed as 
\equ{ 
B_{\ua b}  = \half \big(\bK_{\ua, b} - K_{b, \ua} \big)~, 
\qquad 
B_{a b} = \half \big( K_{a, b} - K_{b, a} \big)~, 
\qquad 
\bB_{\ua\, \ub} = \half\big(\bK_{\ua, \ub} - \bK_{\ub, \ua} \big)~, 
\label{Bfield} 
}
in terms of derivatives of $K_a$ and $\bK_\ua$.

The scalar component representation is not unique, because one may add total derivatives. By different partial integrations of the expression $\int\d^2\gs\, K_a \derl\derr z^a$ one obtains the identity 
\begin{subequations}
\equ{
B_{ab} \derl z^a \derr z^b = \big(B_{\ua a} - G_{\ua a} \big)\,  \frac 12 
\Big( \derl z^a \derr \bz^\ua - \derl \bz^\ua \derr z^a \Big)~.
\label{Brewrite}
}
Similarly from $\int\d^2\gs\, \bK_\ua \derl\derr \bz^\ua$ one finds 
\equ{
\bB_{\ua\,\ub} \derl \bz^\ua \derr \bz^\ub = \big(B_{\ua a} + G_{\ua a} \big)\,  \frac 12 
\Big( \derl z^a \derr \bz^\ua - \derl \bz^\ua \derr z^a \Big)~.
\label{bBrewrite}
}
\end{subequations} 
Using these relations we may remove the $B_2$--components $B_{\ua b}$, i.e.\ set $\gb = 1$, or remove $B_{ab}$ and $\bB_{\ua\,\ub}$, i.e.\ set $\gb = 0$. This reflects different gauge choices for the Kalb--Ramond field, see e.g.~\cite{Nibbelink:2010wm}.  However, it is not possible to remove all $B_2$--field components simultaneously.  In the following we exploit this, by identifying the $B$--field by its (anti--)holomorphic components, $B = [B_{ab}]$, $\bB = [\bB_{\ua\,\ub}]$, only.

Even though the classical component action~\eqref{ScalarAcChiral} depends on the parameter $\gb$, the resulting classical equations of motion for the coordinate fields can be written as 
\begin{subequations}
\label{classicalEoM} 
\equ{
G_{\ub a}\, \derl\derr z^a + G_{\ub a, b}\, \derl z^b \derr z^a
- \bB_{\ua\,\ub,b}\, \derl z^b \derr \bz^\ub = 0~, 
\\[2ex] 
G_{\ua b}\, \derl\derr \bz^\ua + G_{\ua b,\ub}\, \derl\bz^\ub \derr \bz^\ua
- B_{ab,\ub}\, \derl\bz^\ua \derr z^b = 0~, 
}
\end{subequations} 
independently of the $B_2$--field gauge.

The presence of a non--trivial two--form $B_2$ introduces torsion on the complex manifold~\cite{Strominger:1986uh}. In heterotic theories the amount of torsion is measured by the three--form $H_3=dB_2$, see e.g.~\cite{LopesCardoso:2002hd}, and signals that the manifold is non--\Kh. In $(2,0)$ superspace its non--vanishing components are 
 \equ{
 H_{a b \uc} = H_{\uc a b} = H_{b \uc a} = 
 K_{a,b\uc} - K_{b,a\uc}~, 
 \qquad 
 H_{\ua\,\ub c} = H_{c \ua\,\ub} = H_{\ub c \ua} = 
 \bK_{\ua,\ub c} - \bK_{\ub,\ua c}~; 
 \label{compH3} 
 }
 the purely (anti--)holomorphic components vanish identically: $H_{abc}=H_{\ua\,\ub\,\uc}= 0$.

We introduce the torsion connections
\equ{ 
\gG_{\!\pm\,}{}^{b}_{cd} = 
G^{b \ua} \Big( G_{\ua c,d} \pm \half\, H_{cd \ua} \Big)~, 
\qquad 
\bgG_{\!\pm\,}{}^{\ub}_{\uc\,\ud} = 
G^{a \ub} \Big( G_{\uc a,\ud} \pm \half\, H_{\uc\,\ud a} \Big)~, 
\label{TorsionConnections}
}
which generalize the Christoffel connections 
$\gG^{b}_{cd} = G^{b \ua}\, G_{\ua c,d}$ and 
$\bgG^{\ub}_{\uc\,\ud} = G^{a \ub}\, G_{\uc a,\ud}$ 
of a Hermitean manifold. The presence of torsion in the connections $\gG_{\!\pm}$ is made apparent by the fact that they do not posses any  symmetry in its lower indices $c$ and $d$; an anti--symmetric part enters through the components \eqref{compH3} of the three--form $H_3$. The properties of the various torsion connections are collected in Appendix~\ref{sc:Connections}. The kinetic terms for the right--moving fermions $\gps$ select torsion connections, $\gG_{\!+}, \bgG_{\!+}$, as their component action 
\equ{ 
S_\gf \supset  
\frac{i}{2} \int\d^2\gs\, 
\Big\{ 
 \bgps^\ua\, G_{\ua b}\,  
\big(\derr \gps^b \,+\, \gG_{\!+\,}{}^{b}_{cd}\, \derr z^c\, \gps^d\big) 
-  \big( \derr \bgps^\ua \,+\, 
\bgG_{\!+\,}{}^{\ua}_{\uc\,\ud}\, \derr \bz^\uc\, \bgps^\ud \big) 
\, G_{\ua b} \, \gps^b
\Big\}~.
}
can be conveniently written in terms of them only.

\subsubsection*{Holomorphic vector bundle}

The left--moving fermions define a holomorphic vector bundle on the complex manifold. Indeed, the quadratic action for these fermions can be written in the form 
\equ{
S_\gL \supset  
\frac{i}{2} \int\d^2\gs\, 
\Big\{       
 \bgl^{\uga}\, N_{\uga\gb}\, 
\Big( \derl \gl^\gb + [A_c]^\gb{}_\gg\, \derl z^c\, \gl^\gg \Big)
- 
\Big( \derl \bgl^\uga N_{\uga\gb}
+ \bgl^\uga\,N_{\uga\gg} [A_\uc]^\gg{}_\gb\, \derl \bz^\uc \Big) 
\,  \gl^\gb  + 
\non \\[2ex] 
 -  \gl^\ga\, [W_\uc]_{\ga\gb}\, \derl \bz^\uc\,   \gl^\gb   
+   \bgl^\uga\, [\bW_c]_{\uga\,\ugb}\, \derl z^c\,  \bgl^\ugb 
\Big\}~, 
\label{FermiQuadraticAction}
}
where the connections 
\equa{
[A_c]^\gb{}_\gg = N^{\gb\ugg}\,N_{\ugg\gg,c}~,
\quad 
[A_\uc]^\gb{}_\gg  =  N^{\gb\ugg}\, N_{\ugg\gg,\uc}~,
\qquad 
[\bW_c]_{\uga\,\ugb}  = \bM_{\uga\,\ugb,c} ~,
\quad 
[W_\uc]_{\uga\,\ugb} =  M_{\ga\gb,\uc}~.
\label{GaugeConnections}
}
can  be interpreted as target space gauge fields. In these expressions the inverse of the matrix $N$ has been denoted by 
$[N^{-1}]^{\gb\ugg} = N^{\gb\ugg}$ with upper indices. In detail, the first two connections define $U(16)$ gauge fields. The latter two define conjugate vector fields which form rank $(2,0)$ and $(0,2)$ representations of $U(16)$. Hence, in total we have 
$16^2 + 2\cdot \half (16\cdot 15) = 496$ target space gauge fields. In other words, the gauge connections~\eqref{GaugeConnections} correspond to the branching of $SO(32) \ra U(16)$; only the $U(16)$ part is realized linearly, see e.g.~\eqref{gauge} below.

The associated gauge field strengths, $F_{\ua a}(N)$, $F_{\ua a}(M)$ and $F_{\ua a}(\bM)$, can be read off from the four--fermion terms 
\equ{
S_\gL 
 \supset 
\int\d^2\gs\, \Big\{  
\bgl^\uga \, [F_{\uc c}(N)]{}_{\uga\gb} \, \gl^\gb   
+ \half\, \gl^\ga\, [F_{\uc c}(M)]_{\ga \gb} \,\gl^\gb  
+ \half\, \bgl^\uga\,  [F_{\uc c}(\bM)]_{\uga \ugb}\,\bgl^\ugb
\Big\} \bgps^\uc\gps^c~, 
\label{FourFermionTerms} 
}
after the auxiliary fields, $h^\ga$ and $\bh^\uga$, have been eliminated. They can be written as  
\begin{subequations}
 \label{GaugeFieldStrength}
\equ{
F_{\uc c}(N)    =  
N_{,\uc c}  - N_{,\uc} \, N^{-1}\, N_{,c}    
+ \bM_{,c}\, (N^{-1})^T\, M_{,\uc}~, 
\\[1ex] 
F_{\uc c}(M)   = 
M_{,\uc c} - M_{,\uc}\, N^{-1}\, N_{,c}     
+ (N^{-1}\, N_{,c})^T\,  M_{,\uc}~, 
\\[1ex]  
F_{\uc c}(\bM) =  
\bM_{,\uc c}    
- N_{,\uc} \, N^{-1}\, \bM_{,c} 
+ \bM_{,c}\, (N_{,\uc}\, N^{-1})^T ~, 
}
\end{subequations} 
without explicitly indicating their gauge indices $\ga, \uga$, etc.

\subsection{Reparameterizations} 
\label{sc:Reparameterizations}

\subsubsection*{Holomorphic redefinitions}

The functions $K(\gf,\bgf)$ and $\bK(\gf,\bgf)$ in \eqref{ClAction} are defined up to the addition
\equ{
K(\gf,\bgf) \ra K(\gf,\bgf) + k(\gf)~, 
\qquad 
\bK(\gf,\bgf) \ra \bK(\gf,\bgf) + \bk(\bgf)~, 
\label{redefKbK}
}
 of holomorphic $k(\gf)$ and anti--holomorphic $\bk(\bgf)$ functions, respectively, because chiral superfields integrated over full superspace vanish. These transformations leave the metric~\eqref{Metric} and the mixed $B$--field components~\eqref{Bfield} inert. However, the pure (anti--)holomorphic parts of the $B_2$--field components transform as 
 \begin{subequations}
 \label{holoBtrans} 
 \equ{
 B_{ab}(\gf,\bgf)  \ra B_{ab}(\gf,\bgf) 
 + \sfrac 12\, k_{a,b}(\gf) - \sfrac 12\, k_{b,a}(\gf)~, 
 \\[2ex] 
 \bB_{\ua\,\ub}(\gf,\bgf)  \ra \bB_{\ua\,\ub}(\gf,\bgf) 
 + \sfrac 12\, \bk_{\ua,\ub}(\bgf) - \sfrac 12\, \bk_{\ub,\ua}(\bgf)~, 
 }
 \end{subequations} 
Only these (anti--)holomorphic transformations of the $B_2$--field are explicitly visible in the (2,0) superspace formulation.

Similarly, the functions $M(\gf,\bgf)$ and $\bM(\gf,\bgf)$ are defined only up to the addition 
\equ{
M(\gf,\bgf) \ra M(\gf,\bgf) + m(\gf)~, 
\qquad 
\bM(\gf,\bgf) \ra \bM(\gf,\bgf) + \bm(\bgf)~, 
\label{redefMbM} 
}
of holomorphic $m(\gf)$ and anti--holomorphic $\bm(\bgf)$ functions, respectively.

\subsubsection*{Target space diffeomorphisms}

The chiral superfields are defined up to holomorphic reparameterizations 
\equ{ 
\gf \ra f(\gf)~, 
\qquad 
\bgf \ra \bff(\bgf)~, 
\label{diffTarget}
}
in order to preserve the chirality property~\eqref{chiralcond} of $(2,0)$ chiral multiplets. In target space these correspond to holomorphic diffeomorphisms that preserve the chosen complex structure. The classical worldsheet action~\eqref{ClAction} is invariant provided that the functions $N, M$, $\bM$ transform as scalars, e.g.\ 
$N(\gf,\bgf) \ra N(f(\gf), \bff(\bgf))$, and the functions $K$ and $\bK$ as one--form components 
\equ{
K(\gf, \bgf) \ra K(f(\gf), \bff(\bgf))\, X(\gf)~, 
\qquad 
\bK(\gf, \bgf) \ra \bX(\bgf)\, \bK(f(\gf), \bff(\bgf))~, 
\label{diffKbK}
}
where $[X(\gf)]^a{}_b = f^a{}_{,b}(\gf)$ and $[\bX(\bgf)]^\ua{}_\ub = \bff^\ua{}_{,\ub}(\bgf)$. The metric and the anti--symmetric (anti--)holomorphic $B_2$--field components transform as 
\equ{ 
G(\gf,\bgf) \ra \bX(\bgf)\, G(f(\gf), \bff(\bgf))\, X(\gf)~, 
\qquad 
\arry{l}{
B(\gf,\bgf) \ra X^T(\gf)\, B(f(\gf),\bff(\bgf))\, X(\gf)~, 
\\[1ex] 
\bB(\gf,\bgf) \ra \bX(\bgf)\, \bB(f(\gf),\bff(\bgf))\, \bX^T(\bgf)~. 
}
}

\subsubsection*{Gauge transformations}

Similarly, only a part of the target space gauge transformations are visible in the $(2,0)$ superspace description. The chirality constraints~\eqref{fermichiralcond} on the Fermi superfields imply that the allowed redefinitions take the form 
\equ{ 
\gL \ra g(\gf)\, \gL~, 
\qquad 
\bgL \ra \bgL\, \bg(\bgf)~, 
\label{gauge}
}
where $[g(\gf)]^\ga{}_\gb$ and $[\bg(\bgf)]^\ua{}_\ub$ define the matrix components of (anti--)holomorphic $U(16)$ transformations. The matrix functions $N, M$ and $\bM$ consequently transform as 
\equ{ 
N(\gf,\bgf) \ra \bg(\bgf)\, N(\gf,\bgf)\, g(\gf)~, 
\qquad 
\arry{l}{
M(\gf,\bgf) \ra g^T(\bgf)\, M(\gf,\bgf)\, g(\gf)~, 
\\[2ex] 
\bM(\gf,\bgf) \ra \bg(\bgf)\, \bM(\gf,\bgf)\, \bg^T(\bgf)~. 
}
}
Inserting these transformations in the definitions of the target space gauge potentials~\eqref{GaugeConnections} shows that $A_c$ and $A_\uc$ transform as U(16) gauge connections 
\equ{
A_c \ra g^{-1}\big( A_c + \der_c \big) g~, 
\qquad 
A_\uc \ra g^{-1}\big( A_\uc + \bder_\uc \big) g~, 
}
while the other components $W_\uc$ and $\bW_c$ transform as rank--two anti--symmetric tensors, i.e.\ 
$W_\uc \ra g^T\, W_\uc\, g$ 
and 
$\bW_c \ra \bg\, \bW_c\, \bg^T$. Hence the $(2,0)$ description only realizes the $U(16) \subset SO(32)$ gauge group manifestly.\footnote{To see that $W_\uc$ and $\bW_c$ transform as $SO(32)/U(16)$ gauge connections we have to allow for transformations that change the complex structure. As this breaks manifest (2,0) supersymmetry, we do not consider this here.} The gauge field strengths \eqref{GaugeFieldStrength} all transform covariantly under $U(16)$ transformations.

\subsection{Classical Weyl invariance}
\label{sc:superWeyl} 

The worldsheet theory has to be invariant under worldsheet diffeomorphisms and Weyl transformations 
\equ{
\text{\cal g}(\gs) \ra e^{2\, \go(\gs)}\,\text{\cal g}(\gs)~, 
\qquad 
\sqrt{\text{\cal g}} R \ra 
\sqrt{\text{\cal g}} (\text{\cal R} - 2 \nabla^2 \go)~,
\label{Weyl} 
}
of the worldsheet metric $\text{\cal g}(\gs)$ with Ricci scalar $\text{\cal R}$. In most of this work we assume that we have used combined Weyl and worldsheet diffeomorphism to bring the metric to the flat gauge $\text{\cal g}(\gs) = \text{diag}(-1,+1)$. To describe a heterotic worldsheet theory with a generic metric $\text{\cal g}$, we have to consider (1,0) supergravity. By introducing the usual ghost sector implies that the critical dimension of the heterotic string is 10.

The (2,0) theory considered above on a flat worldsheet possesses (2,0) super conformal symmetry. Therefore, one would expect that also the consequences of Weyl invariance can be analyzed in a manifest (2,0) supersymmetric fashion. Only to investigate when such (2,0) super Weyl transformations are symmetries of the (regularized) quantum theory we resort to (2,0) supergravity. To be sure we do not mean to imply that we consider (2,0) supergravity as the fundamental description of the heterotic string\footnote{We are indebted to Ilarion Melnikov and Savdeep Sethi for pointing out that our previous version of the manuscript could be misinterpreted on this point.}, for then the target space dimensions will be 4 with either Euclidean or (2,2) signature~\cite{Ademollo:1976wv} (and e.g.\ Ref.~\cite{Marcus:1992wi} for a modern review). However, the super Weyl invariance will allow us to trace how the dilaton gets involved in a very systematic way in Subsection~\ref{sc:SuperWeylAnom}.

Coupling our NLSM to (2,0) supergravity basically involves the following modifications to the (2,0) NLSM action~\cite{Brooks:1986uh,Evans:1986ada,Brooks:1987nt}: i) Replace the  derivatives, $D_+, \bD_+$ and $\derr$, by diffeormorphism covariant ones, $\nabla_{\!+},\bnabla_{\!+}$ and $\nabla_{\!R}$, respectively. ii) Include the super vielbein measure $\cE$ in the action~\eqref{ClAction}. iii) The (2,0) supergravity action~\cite{Evans:1986ada,Brooks:1987nt}
\equ{
S_\gPs = \frac 1{4\gp} \int\d^2\gs\d^2\gth^+\, \cE\, \gPs(\gf,\bgf)\, \cG_R \supset 
\frac 1{4\gp} \int \d^2\gs\, \sqrt{\text{\cal g}}\, \gPs(z,\bz)\, \text{\cal R}~. 
\label{DilatonAction} 
}
has to be added~\cite{Fradkin:1985fq}. The dilaton $\gPs(\gf,\bgf)$ is a general real function of the chiral superfields and their conjugates. Expanding this action in components gives the usual coupling of the dilaton to the worldsheet Ricci curvature scalar $\text{\cal R} = \frac 12[D_+, \bD_+] \cG_R|$.

In a (2,0) theory the Weyl transformation~\eqref{Weyl} is replaced by a super Weyl transformation, which infinitesimally acts as~\cite{Evans:1986wp,Brooks:1987nt,Govindarajan:1991sx}
\equ{
\arry{c}{
\gd_\cS \gf = 0~, 
\qquad 
\gd_\cS (\nabla_{\!R\,} \gf) = - \cS\, \nabla_{\!R} \gf~, 
\qquad 
\gd_\cS \gL = - \sfrac 12\, \cS\, \gL~, 
\\[2ex] 
\gd_\cS \cE = \cS\, \cE~, 
\qquad 
\gd_\cS \cG_R = -\cS\, \cG_R  + \nabla_{\!R\,} \cU~, 
}
\label{superWeyl} 
}
where the super Weyl parameters $\cS$ and $\cU$  are a real (2,0) superfields satisfying the constraints~\cite{Evans:1986wp}
\equ{
\nabla_{\!+\,} \cS = i\, \nabla_{\!+\,} \cU~, 
\qquad 
\bnabla_{\!+\,} \cS = -i\, \bnabla_{\!+\,} \cU~, 
\label{SUconstr}
}
with $S| = 2 \go$. Hence, $\cS$ and $\cU$ can be thought of as the real and imaginary part of a chiral superfield. Using combined super Weyl transformations and super diffeormorphisms we can make $\cG_R=0$ locally.  Classically the super Einstein--Hilbert action~\eqref{DilatonAction} is only super Weyl invariant, when the dilaton $\gPs$ is constant. In Subsection~\ref{sc:SuperWeylAnom} we will see that the super Weyl anomaly forces the dilaton to be non--trivial in general.

\section{Quantum (2,0) non--linear sigma models}
\label{sc:Quantum}

We can compute the quantum corrections to the classical action as an expansion in loop diagrams. To setup this analysis we have to define the effective quantum action. To compute loop diagrams we need to identify the propagators and the relevant vertices. We present some special contraction properties of supergraphs which prove very useful in computing quantum corrections at the end of this Section.

\subsection{Effective action}
\label{sc:EffectiveAction}

To define the effective action we consider the expansion of the theory around some classical background, denoted collectively by  $\gx = (\gf, \bgf, \gL, \bgL)$. The full quantum superfields $\gx_\text{full} = \gx + \gp(\gX)$ are decomposed into the background fields $\gx$  and the quantum superfields $\gX$. Define the generating functional as the path integral 
\equ{
e^{i W(\gx, \gTh)} = \int \cD \gX\,  
\exp \Big\{ i\, S\big(\gx + \gp(\gX)\big) + i\, \gX\cdot \gTh \Big\}~,  
}
with superfield sources $\gTh$. The classical / quantum splitting, encoded in the function $\gp$, is assumed to be chosen such that the quantum fields $\gX$ transform covariantly. This ensures that the functional $W(\gx,\gTh)$ formally possesses the same symmetries as the classical action $S(\gx)$. By the Legendre transform 
\equ{
\gG(\gx, \gX_0) = W(\gx, \gTh) - \gTh\cdot \gX_0~, 
}
we obtain an effective action $\gG(\gx, \gX_0)$ in which the sources $\gTh$ have been replaced by mean fields $\gX_0$. By functional differentiating the Legendre transform one obtains an expression of the sources $\gTh$. After a shift $\gX \ra \gX_0 + \gX$ of the integration variables, the effective action may be written as~\cite{Howe:1986vm,Hull:1986hn}
\equ{ 
e^{i\gG(\gx,\gX_0)} = \int \cD\gX\, 
\exp \Big\{ i\, S\big(\gx + \gp(\gX_0+\gX)\big) 
- i\, \frac{\gd \gG}{\gd \gX_0}(\gx,\gX_0) \cdot \gX \Big\}~. 
}
When the classical / quantum splitting is linear, the background fields $\gx$ and the mean fields $\gX_0$ only appear in the combination $\gx+\gX_0$, hence their roles can essentially be interchanged. In general, we would like that the effective action can be thought of as an action that takes quantum effects into account. Since the classical action $S(\gx)$ is a function of $\gx$ only, this motivates to define the effective action $\gG(\gx) = \gG(\gx, 0)$ by simply setting $\gX_0=0$ in the expression above: 
\equ{
e^{i\gG(\gx)} = \int \cD\gX\, 
\exp \Big\{ i\, S\big(\gx + \gp(\gX)\big) 
- i\, \frac{\gd \gG}{\gd \gX_0}(\gx,0) \cdot \gX \Big\}~. 
\label{EffectiveAction}
}
The role of the second term in the exponential is to ensure that this is the  one--particle--irreducible (1PI) action w.r.t.\ the classical / quantum splitting defined by $\gp$.

The requirement that the quantum superfields $\gX$ transform covariantly strongly restricts the form of the function $\gp$ that encodes the classical / quantum splitting. In particular, this decides whether a linear splitting is sufficient or a non--linear splitting is necessary. Concretely, the quantum superfields $\gX$ have to transform as covariant vectors w.r.t.\ holomorphic target space diffeomorphisms~\eqref{diffTarget} and gauge transformations~\eqref{gauge}. This implies that for the superfields $\gL, \bgL$ the linear splitting is sufficient, because they transform linearly under the gauge transformations~\eqref{gauge}.

For the superfields $\gf, \bgf$ the situation is more complicated, because their scalar components $z,\bz$  take values on a complex torsion manifold as discussed in Subsection~\ref{sc:action}. As the full quantum superfield $\gf_\text{full} = \gf + \gp(\gF)$ transform according to~\eqref{diffTarget} as dictated by target space diffeomorphisms, $\gF$ has to transform as a tangent space vector. To ensure this one typically uses Riemann normal coordinates to set up the definition of the effective action~\cite{Hull:1986hn}. However, in the present case the field redefinition $\gf_\text{full} = \gf+ \gp(\gF)$ needs to be holomorphic because of the chirality condition~\eqref{chiralcond}. An extension of normal coordinates to K\"ahler manifolds has been proposed in~\cite{Higashijima:2000wz,Higashijima:2002fq}. Unfortunately, the use of such normal coordinates does not seem to be an option in our context: In general a (2,0) NLSM possesses torsion~\eqref{compH3} which renders the normal coordinate expansion non--holomorphic and thus incompatible with the (2,0) chirality conditions~\eqref{chiralcond} of the coordinate superfields $\gf$. (As discussed in~\cite{Hull:1985zy,Hull:1986hn} a similar issue arrises for (2,1) supersymmetry.)  For this reason we directly define holomorphic normal coordinates below, which preserve holomorphic general covariance while being compatible with (2,0) supersymmetry.

Holomorphic normal coordinates can be thought of as Riemann normal coordinates for complex manifolds using purely holomorphic coordinates instead of real ones. To define the holomorphic normal coordinates, we observe that we can obtain a holomorphic geodesic equation 
\equ{
\ddot \gf^a(t) + \tgG^a_{bc}\big(\gf(t), \bgf\big)\, \dot\gf^b(t)\, \dot\gf^c(t) = 0~, 
\label{geodesic}
}
from the classical equations of motion~\eqref{classicalEoM} for a curve defined by $\gf(t)$ for $0 \leq t \leq 1$, while keeping $\bgf(t) = \bgf$ fixed. Here the connection $\tgG$ is the unique symmetric Hermitean connection~\eqref{symmConnection} defined in Appendix~\ref{sc:Connections}: As the torsion~\eqref{compH3} involve mixed indices, it does not appear in the holomorphic geodesic equation. Moreover, the torsion connection which is present in the classical equations of motion~\eqref{classicalEoM} gets symmetrized in its indices $b,c$ by the contraction with $\dot\gf^b(t)$ and $\dot\gf^c(t)$. By taking the boundary conditions, $\gf(0) = \gf$,  $\gf(1) = \gf + \gp(\gF)$, and defining $\gF = \dot \gf(0)$, we find that 
\equ{
\gf(1) = \gf + \gF + \sum_{n\geq 2}\, \frac 1{n!}\, \gf^{(n)}(0)~.
}
Using the holomorphic geodesic equation~\eqref{geodesic} the higher order derivatives $\gf^{(n)}(0)$ can be determined recursively. In this work we only need the first few orders 
\equ{
\gp^a(\gF) = \gF^a - \frac 12\, \tgG^a_{bc} \gF^b \gF^c +\ldots 
\label{HolomorphicNormalCoordinates}
}
Since $\gF$ is defined as a derivative, it transforms covariantly under holomorphic coordinate transformations.  As far as the path integral measure is concerned this defines a holomorphic transformation as only $\gf_\text{full}$ and $\gF$ are integration variables.

\subsection{Super propagators}
\label{sc:Propagators}

The super propagators can be read off from the quadratic part of the action expanded to second order in the quantum superfields, see e.g.~\cite{Brooks:1986gd,Govindarajan:1991sx,Cui:2010si,Cui:2011rz}. Even though this procedure is in principle standard, we should point out some features that are specific to our work: Since we aim only to determine the beta functions of the (2,0) theory, we are not interested in the full quantum effective action but only in the renormalization of the terms of the classical action. Given that they all are only functions of the chiral superfields $\gf,\bgf$, we may essentially assume that in the loop computations the expansion coefficients are all mere constants rather than superfield expressions.

\subsubsection*{Fermi super propagator}

The coupling of the sources $J, \bJ$ to the Fermi superfields $\gL, \bgL$ is described by the action 
\equ{
S_J = \int \d^2\gs 
\Big\{
\int\d\gth^+\, J^T \gL + \int\d\bgth^+\, \bJ\, \bgL^T
\Big\}~. 
\label{FermiSources} 
}
Consequently, the source $J$ is a bosonic chiral superfield. Because $\gL$ is fermionic, the functional differentiation $\fd{J}$ is also fermionic. This is consistent with the definition of functional differentiation w.r.t.\ to this source 
\equ{
\fd[J_{\gb 2}]{J_{\ga 1}}  = \gd^\ga_\gb\, \bD_{+2} \gd_{21}~.
}
As explained in Appendix~\ref{sc:Conventions} we use a subscript to indicate in which coordinate system a certain superfield or operator is evaluated. The definition of the superspace delta function $\gd_{21}$ is also given in Appendix~\ref{sc:Conventions}. Since the full superspace $\gd_{21}$ is bosonic, the super covariant derivative $\bD_{+2}$ makes the expression fermionic. By functional differentiation of the exponentiated source action we can bring factors of $\gL$ down 
\equ{ 
\gL_1^\ga \, e^{i S_J} =  i \fd{J_{\ga 1}}\, e^{i S_J}~. 
}

As the classical action is quadratic in $\gL, \bgL$ we may take the coefficient functions $N, M$ and $\bM$ as constants. Under this assumption the terms with $M$ (and $\bM$) are purely (anti--)holomorphic, and hence irrelevant for the Fermi superfield propagator. Consequently, it is given by 
\equ{ 
iS_{\gL}(J) =  \int\d^2\gs\d^2\gth^+\, \bJ \, (N^T)^{-1} \frac 1{\derl}\,J~. 
\label{FermiProp} 
} 
To obtain this expression we have used that on a chiral superfield the combination of super covariant derivatives, $\bD_+D_+$, can be replaced by a worldsheet derivative $\derl$, see~\eqref{bDDonChiral}. As indicated in the next equation we graphically represent the Fermi superfield by a solid line with an arrow. The arrow defines the flow of chirality determined by the order of the super covariant derivatives that arise by the functional differentiation: 
\equ{ 
\raisebox{-2.5ex}{
\mbox{$
 \begin{fmfgraph*}(20,10) \fmfkeep{FermiPropBetweenVertices}
 \fmfleft{i} 
 \fmfright{o} 
 \fmf{fermion}{i,o} 
 \fmfdot{i,o} 
 \fmflabel{$1$}{i}
 \fmflabel{$2$}{o}
 \end{fmfgraph*}
$}} 
\qquad:\qquad 
 i\fd{ J_2}\, i\fd{\bJ_1} \, iS_\gL(J) = (N^T)^{-1}\, 
\Big( \frac{\bD_+ D_+\, \derr}{-\Box_D+m^2} \Big)_2 \gd_{12}~. 
\label{FermiPropBetweenVertices}
} 
Here we have replaced the naive propagator by  the dimensional regularized one defined in Appendix~\ref{sc:DimRed}. In particular, $\Box_D$ is the d'Alembertian in $D$ dimensions and $m$ denotes an infrared regulator mass.

\subsubsection*{Chiral super propagator}

The computation of the chiral super propagator is a bit more involved,  because the expansion to second order in the quantum superfields contains terms in which the right--moving derivative $\derr$ acts on a quantum superfield or on a background superfield. Moreover, when the right--moving derivative acts on a quantum superfield, one can perform a partial integration to have it act on the other quantum field or on the background function. We choose to perform the partial integrations in a symmetric fashion so as to obtain results which can be expressed only in terms of the background metric $G_{\ua a}$ with mixed indices~\eqref{Metric} and the background Kalb--Ramond $B_2$--field~\eqref{Bfield}  with purely (anti--)holomorphic indices $B_{ab}$ ($\bB_{\ua\,\ub}$). These manipulations result in the expression
\equ{
S_\gF = \frac i4 \int\d^2\gs \d^2\gth^+\, 
\Big\{  
\derr \bgF^\ua  
\, G_{\ua a} \, \gF^a 
- 
\bgF^\ua \,G_{\ua a}\,  
\derr \gF^a 
\Big\}~;
\label{QuadrQuantChiral} 
}
where terms involving $B_{ab}$ and $\bB_{\ua\,\ub}$ have been dropped as they do never give contributions in diagrams by chirality.

The sources $\gO, \bgO$ for the chiral superfields $\gF, \bgF$ are Fermi superfields to ensure that the source action, 
\equ{
S_\gO = \int \d^2\gs 
\Big\{
\int\d\gth^+\, \gO^T \gF + \int\d\bgth^+\, \bgO\, \bgF^T
\Big\}~,
\label{ChiralSources} 
}
is a c--number. Consistency of the definition of functional differentiation $\fd{\gO}$ requires that it is bosonic: 
\equ{
\fd[\gO_{b 2}]{\gO_{a 1}}  = \gd^\ga_\gb\, \bD_{+2} \gd_{21}~. 
}
The functional differentiation of the exponentiated source action brings factors of $\gF$ down 
\equ{ 
\gF_1^a \, e^{i S_\gO} =  \fd{i \gO_{a 1}}\, e^{i S_\gO}~. 
}

The terms in the action~\eqref{QuadrQuantChiral} in which the right--moving derivative $\derr$ acts on the quantum superfields $\gF, \bgF$ can be used to derive their propagator
\equ{ 
iS_{\gF}(\gO) =  \int\d^2\gs\d^2\gth^+\, \bgO \, (G^T)^{-1} \frac {-i}{\derl\derr}\, \gO~. 
\label{ChiralProp} 
} 
The chirality that propagates between two vertices is again determined by the order of the super covariant derivatives like in~\eqref{FermiPropBetweenVertices}: 
\begin{subequations}
\label{ChiralProps} 
\equa{ 
\raisebox{-2.5ex}{
\mbox{$
 \begin{fmfgraph*}(20,10) \fmfkeep{ChiralPropBetweenVertices}
 \fmfleft{i} 
 \fmfright{o} 
 \fmf{scalar}{i,o} 
  \fmfv{decor.shape=cross,decor.size=4thick}{i,o} 
 \fmflabel{$1$}{i}
 \fmflabel{$2$}{o}
 \end{fmfgraph*}
$}} 
\qquad & : \qquad 
 \fd{ i\gO_1}\, \fd{i \bgO_2}\, i S_\gF(\gO) = -i \, (G^T)^{-1}\, 
\Big[ \frac{\bD_+ D_+}{-\Box_D+m^2} \Big]_2 \gd_{12}~. 
\label{ChiralPropBetweenVertices} \\ 
%
\intertext{It sometimes happens that a $\derr$ derivative acts on a chiral superfield line. This changes a chiral propagator effectively into a Fermi propagator, to distinguish it from the true Fermi propagator~\eqref{FermiPropBetweenVertices}, we draw crosses at its vertices: 
}
%
\raisebox{-2.5ex}{
\mbox{$
 \begin{fmfgraph*}(20,10) \fmfkeep{ChiralPropBetweenVertices1derr}
 \fmfleft{i} 
 \fmfright{o} 
 \fmf{fermion}{i,o} 
  \fmfv{decor.shape=cross,decor.size=4thick}{i,o} 
 \fmflabel{$1$}{i}
 \fmflabel{$2$}{o}
 \end{fmfgraph*}
$}} 
\qquad & : \qquad 
 \fd{ i\gO_1}\, \derr\, \fd{i \bgO_2}\, i S_\gF(\gO) = -i \, (G^T)^{-1}\, 
\Big[ \frac{\bD_+ D_+\,\derr}{-\Box_D+m^2} \Big]_2 \gd_{12}~. 
\label{ChiralPropBetweenVertices1derr}
%
\intertext{Since a propagators sits between two vertices, it might even happen that both vertex derivatives act on the same chiral propagator. This case we denote by a double line between the chiral vertices: }
%
\raisebox{-2.5ex}{
\mbox{$
 \begin{fmfgraph*}(20,10) \fmfkeep{ChiralPropBetweenVertices2derr}
 \fmfleft{i} 
 \fmfright{o} 
 \fmf{heavy}{i,o} 
  \fmfv{decor.shape=cross,decor.size=4thick}{i,o} 
 \fmflabel{$1$}{i}
 \fmflabel{$2$}{o}
 \end{fmfgraph*}
$}} 
\qquad & : \qquad 
\derr\, \fd{ i\gO_1}\, \derr \fd{i \bgO_2}\, i S_\gF(\gO) =
i \, (G^T)^{-1}\, 
\Big[ \frac{\bD_+ D_+\, \derr^2}{-\Box_D+m^2} \Big]_2 \gd_{12}~. 
\label{ChiralPropBetweenVertices2derr}
} 
\end{subequations} 
The sign of this expression is opposite to the ones above because one of the $\derr$ derivatives acts on the source in coordinate system $1$; when we change it to system $2$, we pick up a sign.

\subsection{Super vertices}
\label{sc:Vertices} 

\begin{figure}[t]
\begin{center}
\tabu{cccc}{
\qquad 
 \begin{fmfgraph*}(30,20)
 \fmftop{t}
 \fmfbottom{i,o}
 \fmfdot{v}
 \fmf{fermion}{i,v}
 \fmf{fermion}{v,o} 
 \fmf{scalar}{v,t}
 \end{fmfgraph*}
\qquad & \qquad\qquad 
 \begin{fmfgraph*}(30,20) 
 \fmftop{t}
 \fmfbottom{i,o} 
 \fmfdot{v}
 \fmf{fermion}{i,v}
 \fmf{fermion}{v,o} 
 \fmf{scalar}{t,v}
 \end{fmfgraph*}
\qquad & \qquad\qquad  
 \begin{fmfgraph*}(30,20) 
 \fmftop{t1,t2}
 \fmfbottom{b1,b2} 
 \fmfdot{v}
 \fmf{fermion}{b1,v}
 \fmf{fermion}{v,b2} 
 \fmf{scalar}{t1,v}
  \fmf{scalar}{v,t2}
\end{fmfgraph*}
}
 \\[1ex]  
 \tabu{cccc}{
 \begin{fmfgraph*}(30,20)
 \fmftop{t}
 \fmfbottom{i,o}
 \fmfdot{v}
 \fmf{fermion}{i,v}
 \fmf{fermion}{o,v} 
 \fmf{scalar}{v,t}
 \end{fmfgraph*}
\qquad & \qquad
 \begin{fmfgraph*}(30,20) 
  \fmftop{t1,t2}
 \fmfbottom{b1,b2} 
 \fmfdot{v}
 \fmf{fermion}{b1,v}
 \fmf{fermion}{b2,v} 
 \fmf{scalar}{t1,v}
  \fmf{scalar}{v,t2}
\end{fmfgraph*}
\qquad & \qquad 
 \begin{fmfgraph*}(30,20) 
 \fmftop{t}
 \fmfbottom{i,o} 
 \fmfdot{v}
 \fmf{fermion}{v,i}
 \fmf{fermion}{v,o} 
 \fmf{scalar}{t,v}
 \end{fmfgraph*}
\qquad & \qquad 
 \begin{fmfgraph*}(30,20) \fmfkeep{Fermi02Vert}
 \fmftop{t1,t2}
 \fmfbottom{b1,b2} 
 \fmfdot{v}
 \fmf{fermion}{v,b1}
 \fmf{fermion}{v,b2} 
 \fmf{scalar}{t1,v}
  \fmf{scalar}{v,t2}
\end{fmfgraph*}
\\[-3ex] 
 }
  \end{center}
\caption{This Figure collects the relevant three and four point vertices which involve two Fermi superfields.
\label{fg:FermiVertices}} 
\end{figure}
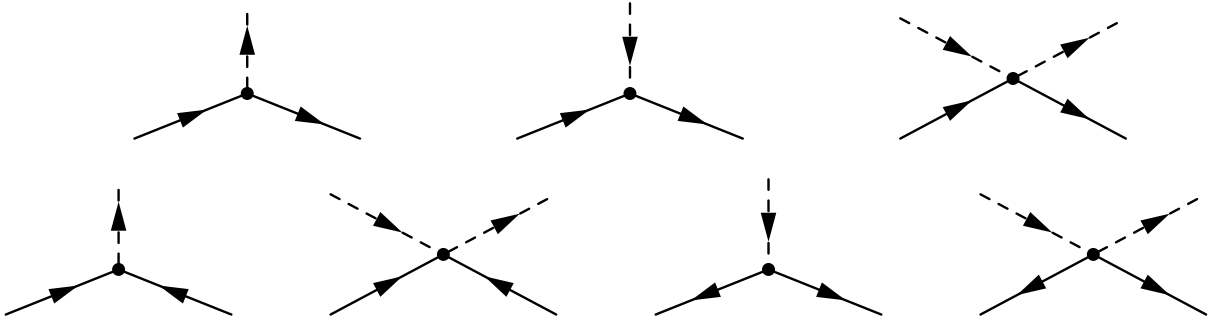

\subsubsection*{Vertices with Fermi superfields}

The vertices involving two Fermi superfields and a number of quantum (anti--)chiral superfields can be obtained by expanding the Fermi action~\eqref{FermiAction}. In this work we only need the following interactions
\equa{ 
i \gD S_\gL  \supset   
\frac {-i}{2} \int & \d^2\gs\d^2\gth^+\, 
\Big\{
\bgL^\uga \gL^\ga 
\Big(
N_{\uga\ga,a}\, \gF^a + 
N_{\uga\ga,\ua}\, \bgF^\ua + 
N_{\uga\ga,\ua a}\, \bgF{}^\ua \gF^a 
\Big) 
\label{FermiVertices} \\[2ex] 
& + \frac 12\, \gL^\ga \gL^\gb
\Big( 
M_{\ga\gb,\ua}\, \bgF^\ua + 
M_{\ga\gb,\ua a}\, \bgF{}^\ua \gF^a 
\Big) 
 + \frac 12\, \bgL^\uga \bgL^\ugb
 \Big( 
\bM_{\uga\,\ugb,a}\, \gF^a + 
\bM_{\uga\,\ugb,\ua a}\, \bgF{}^\ua \gF^a 
\Big)\Big\}~,
\non 
} 
which are represented graphically in Figure \ref{fg:FermiVertices}.

\subsubsection*{Pure chiral superfield vertices}

\begin{figure}[t]
\centering
\tabu{c}{
\tabu{cccc}{
 \begin{fmfgraph*}(30,20) 
 \fmfleft{i}  
 \fmfright{o} 
 \fmfbottom{b} 
 \fmfv{decor.shape=cross,decor.size=4thick}{v}
 \fmf{scalar}{i,v}
  \fmf{scalar}{v,o}
 \fmf{fermion}{v,b} 
 \fmfv{decor.shape=cross,decor.size=4thick}{b}
 \end{fmfgraph*}
 \quad & \quad  
 \begin{fmfgraph*}(30,20) 
 \fmftop{i,o} 
 \fmfbottom{b} 
 \fmfv{decor.shape=cross,decor.size=4thick}{v}
 \fmf{scalar}{i,v}
  \fmf{scalar}{v,o}
 \fmf{fermion}{v,b} 
 \end{fmfgraph*}
\qquad & \qquad  
  \begin{fmfgraph*}(30,20) 
 \fmfleft{i}  
 \fmfright{o} 
 \fmfbottom{b} 
 \fmfv{decor.shape=cross,decor.size=4thick}{v}
 \fmf{scalar}{i,v}
  \fmf{scalar}{v,o}
 \fmf{fermion}{b,v} 
 \fmfv{decor.shape=cross,decor.size=4thick}{b}
 \end{fmfgraph*}
 \quad & \quad  
  \begin{fmfgraph*}(30,20)
 \fmftop{i,o} 
 \fmfbottom{b} 
 \fmfv{decor.shape=cross,decor.size=4thick}{v}
 \fmf{scalar}{i,v}
  \fmf{scalar}{v,o}
 \fmf{fermion}{b,v} 
 \end{fmfgraph*}
}
\\[0ex] 
  }
\caption{The vertices involve up to three quantum chiral superfields corresponding to~\eqref{ChiralVertices}. 
\label{fg:ChiralVertices}} 
\end{figure}
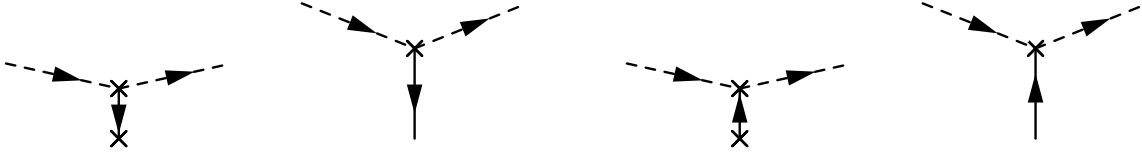

For the computations in the remainder of this paper we need interactions up to cubic order in the quantum chiral superfields.  Some vertices that arise from the kinetic action~\eqref{ChiralAction} contain a single $\derr \gf$ or $\derr \bgf$ derivative, while the others involve three quantum chiral superfields $\gF$ and $\bgF$. These interaction terms read  
\equ{ 
i\gD S_{\gF} \supset  - \frac 14 \int\d^2\gs\d^2\gth^+\Big\{ 
 \bgG_{\! +}{}^\uc_{\ub\,\ua} G_{\uc a}\, \derr \bgf^\ub 
- G_{\ua c} \gG_{\! +}{}^c_{ba}\, \derr \gf^b 
+ \frac 12\, \bH_{\ub\,\ua a}\, \derr \bgF^\ub 
- \frac 12\, H_{ba\ua}\, \derr \gF^b 
\Big\}\, \bgF{}^\ua \,\gF^a~.
\label{ChiralVertices}
} 
The corresponding vertices are depicted in Figure~\ref{fg:ChiralVertices}. The first two interactions are independent of whether normal coordinates or a linear classical / quantum split has been employed. The form of the latter two three--point--interactions crucially depends on the use of the holomorphic normal coordinate expansion~\eqref{HolomorphicNormalCoordinates}.

\subsection{Contraction of supergraphs}
\label{sc:Contraction}

Many supergraphs vanish identically. An often occurring reason for this is that the (contracted) supergraph contains an one--loop Fermi tadpole. It is irrelevant whether it is a fundamental Fermi tadpole or a tadpole of a chiral line with an $\derr$ acting on it. Because the Fermi propagator involves a single $\derr$, such a tadpole is proportional to the integral, 
\equ{ 
\raisebox{-4ex}{
 \begin{fmfgraph}(24,10) 
 \fmfleft{i} 
 \fmfright{o} 
 \fmf{scalar}{i,v1} 
 \fmf{scalar}{v1,o}
 \fmfdot{v1} 
\fmf{fermion,tension=.9}{v1,v1} 
 \end{fmfgraph}
}
\quad\sim\quad 
\int \frac{\d^D q}{(2\gp)^D}\, \frac {q_R}{q^2 + m^2} =0~, 
}
in dimensional regularization, see Appendix~\ref{sc:DimRed}. A dimensional regularized integral is reflection symmetric in their integration variables. But given that this integrand is odd, the reflection symmetry implies that this tadpole has to vanish.

Moreover, as was observed e.g.\ in~\cite{Nibbelink:2005wc,Deguchi:2011rh} in the context $\cN=1$ supergraphs in four dimensions, chiral superfield lines can often be contracted merging two adjacent vertices. Below we show that a similar result naively holds for Fermi superfield lines (or chiral lines with a $\derr$ derivative) in (2,0) theories, but that it is violated in dimensional regularization. However, when the naively contracted diagram is still divergent, the correction due to dimensional regularization can essentially be ignored as far as the determination of counter terms is concerned. On the contrary, when the contracted supergraph would be zero, the dimensional regularization leads to a finite effect which may be interpreted as an anomaly.

\begin{figure}[t]
\centering
\tabu{cccccc}{
\qquad\qquad\qquad  & 
 \begin{fmfgraph*}(40,30) 
 \fmfleft{i1,i2,i3} 
 \fmfright{o}
 \fmfforce{.4w,.5h}{v1}
 \fmfdot{v1,o} 
 \fmf{dbl_dots_arrow}{i1,v1}
 \fmf{dbl_dots_arrow}{i2,v1}
 \fmf{dbl_dots_arrow}{i3,v1} 
 \fmf{fermion,label=$1\hspace{21\unitlength}2$,label.side=right}{v1,o}
 \fmf{dots,left=.1}{i1,i2,i3} 
 \fmflabel{3}{i1}
 \fmflabel{$j$}{i2} 
 \fmflabel{$k$}{i3} 
 \end{fmfgraph*}
&\qquad\qquad& 
\raisebox{10ex}{ 
\tabu{c}{Fermi line \\ contraction \\ ${\longrightarrow}$}}
&\qquad\qquad &
 \begin{fmfgraph*}(40,30) 
 \fmfleft{i1,i2,i3} 
 \fmfright{o}
 \fmfforce{.4w,.5h}{v1}
 \fmfdot{v1,v2} 
 \fmf{dbl_dots_arrow}{i1,v1}
 \fmf{dbl_dots_arrow}{i2,v1}
 \fmf{dbl_dots_arrow}{i3,v1} 
 \fmf{dots,label=$1\hspace{\unitlength}2$,label.side=right}{v1,v2}
 \fmffixed{10,0}{v1,v2}
 \fmf{phantom}{v2,o}
 \fmf{dots,left=.1}{i1,i2,i3} 
\fmflabel{3}{i1}
 \fmflabel{$j$}{i2} 
 \fmflabel{$k$}{i3} 
  \end{fmfgraph*}
  \\[3ex] 
 }
\caption{The contraction of the Fermi propagator relies on the fact that the chiralities are pointing in the same direction.}
\label{fg:FermiContr} 
\end{figure}
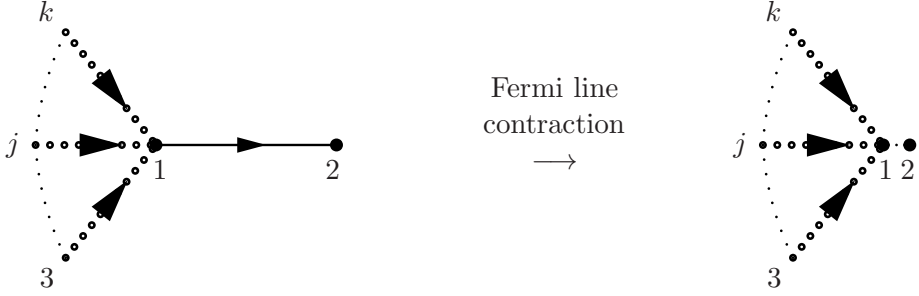

A Fermi super propagator with a chirality pointing in a certain direction can be naively contracted if at one of its vertices all the attached lines have their chiralities aligned in the same direction. In the left picture of Figure \ref{fg:FermiContr} we have sketched such an initial configuration. The solid Fermi propagator extends between the vertices $1$ and $2$. The dotted lines with arrows correspond to any chiral supergraph lines, e.g.\ chiral or Fermi propagators. The collapse of this supergraph comes about for the following reason: As noted in~\eqref{FermiPropBetweenVertices} and~\eqref{ChiralPropBetweenVertices}, the direction of the chirality manifests itself in the order of the super covariant derivatives. This restricts the possibility of partial integrating the super covariant derivatives onto the other lines. So if we can pick any chiral line $j$ in the left picture of Figure \ref{fg:FermiContr} and partially integrate its $\bD_+$, we obtain:  
\equ{
\Big[ \prod_k 
[\bD_+ D_+ ]_1\gd_{k1} \Big]\, 
\Big[ \frac{\bD_+ D_+\, \derr}{-\Box_D+m^2} \Big]_2\,  \gd_{12} 
= - (D_+ )_1 \gd_{j1} \, 
\Big[ \prod_{k\neq j} [\bD_+ D_+ ]_1\gd_{k1} \Big] \, 
\Big[\frac{\bD_+ D_+ \bD_+\, \derr}{-\Box_D+m^2} \Big]_2\ \gd_{12} 
= 
\non \\[1ex]  
=  \Big[ \prod_k [ \bD_+ D_+ ]_1 \gd_{k1} \Big]\, 
\Big[ \frac{2i\, \derl\derr}{-\Box_D+m^2} \Big]_2 \gd_{12} 
=  2i\, \Big[ \prod_k [ \bD_+ D_+ ]_1 \gd_{k1} \Big]\, 
\Big[ 1 + 
 \frac{\der_\perp^2 - m^2}{-\Box_D+m^2} \Big]_2 \gd_{12}~. 
\label{FermiContrPartialIntegration}
}
Because the chirality of all the other lines, $k\neq j$, are pointing in the same direction as that of line $j$, i.e.\ they have the same order of the super covariant derivatives, we get vanishing contributions when we partially integrate $\bD_+$ on them, as super covariant derivatives square to zero~\eqref{AlgebraD}. Therefore, we can only partially integrate its $\bD_+$ derivative on the Fermi propagator. Next, we made use of~\eqref{ChangeOfCoords} to change the coordinate system of the super covariant derivatives. Using~\eqref{AlgebraD} a left--moving derivative $\derl$ appears in the numerator. After this we can partially integrate $\bD_+$ back to its original position.

Then we arrive at the last and essential step: In the non--regularized theory the $\derl\derr$ in the numerator would cancel against the same factor in the denumerator. In the dimensional regularized computation we see that this cancellation still takes place but with a correction term proportional to the momentum in the extra dimensional regularized directions and the IR regulator mass.

\section{Renormalization}
\label{sc:Renorm}

In this section we study the renormalization of the (2,0) NLSM defined in Section~\ref{sc:Classical} and the consequences of super Weyl invariance at the quantum level. To be able to give a systematic treatment of the renormalized theory, we employ dimensional regularization~\cite{'tHooft:1972fi,Collins:1984xc} to regularize the divergent integrals to $D=2-2\ge$ dimensions. The specific adaptation of this scheme employed in this work is described in Appendix~\ref{sc:DimRed}.

\subsection{Beta functions}
\label{sc:BetaFunctions}

Given that the classical action~\eqref{ClAction} is the most general (2,0) action, the counter terms can be compactly parameterized as 
\equ{
\gD S = \int  \d^D\gs \d^2 \gth^+\, \gm^{2\ge} \Big\{
\frac{i}{4} \Big( \derr\bgf\, T(\bK) -T(K)\,\derr\gf \Big)  
- \frac{1}{2}\, \bgL\, T(N)\, \gL
-\frac{1}{4} \Big(  \gL^T \, T(M)\,\gL
+ \text{h.c.} 
\Big) 
\Big\}~, 
\label{CounterTerms}
}
The functions $T(H)$, $H = (H_i) = (K,\bK, N, M, \bM)$ are defined as an expansion in the number of loops $L$ and poles $1/\ge^n$ with $1 \leq n \leq L$,
\equ{
T(H) = \sum_{n \geq 1}\, \frac 1{\ge^n}\, T_{(n)}(H)~, 
\qquad 
T_{(n)}(H) = \sum_{L \geq n} T_{(n,L)}(H)~. 
}
Hence the single pole counter terms, $T_{(1)}(H)$, receive contributions at every loop order in principle. Moreover, all these functions depend on the superfields $\gf,\bgf$ through the functions $H_i = K, \bK, N, M$ and $\bM$ only. This means that the computation of the beta functions in superspace mimics to a large extend the computation of the effective potential in a regular quantum field theory~\cite{Sher:1988mj}; only special care is needed to take the linear term in the $\derr$ derivatives into account. To compute the beta functions we may assume that the expansion coefficients are only functions of $\gs^R$ and otherwise mere constants in superspace. Moreover, as soon as one $\derr$ derivative hits a $\gf$ or $\bgf$, we can assume that all others are strictly constant.

Renormalizability of the (2,0) model implies that the original classical action can be interpreted as defining the bare parameters, $H_{(0)i}$, of the full quantum theory~\cite{Friedan:1980jf,Friedan:1980jm} 
\equ{
H_{(0)} = \gm^{2\ge}\, \Big(
H(\gm) + \sum_{n\geq 1}\, \frac 1{\ge^n}\, T_{(n)}(H)  \Big)~,
}
where the renormalized coupling functions $H(\gm)$ depend on the renormalization scale $\gm$. To trace their scale dependence the beta functions $\gb(H)$ are introduced as 
\equ{
- \ge\, H + \gb(H) = \gb_\ge(H) = \frac 12\, \gm \, \pp{\gm} H(\gm)~. 
\label{defBeta}
}
Renormalizability ensures that $\gb_\ge(H)$ is finite in the limit where the regulator is removed $\ge \ra 0$. The power series expansions in $\ge$ of the bare functions $H_{(0)i}$, which are scale independent, 
\equ{
\ge\, H_{(0)} + \pp[H_{(0)}]{H}\cdot \gb_\ge(H) = \frac 12\,  \gm \, \dd{\gm} H_{(0)} = 0~, 
\label{NoScaleH0}
}
lead to several consequences~\cite{'tHooft:1973mm} (see e.g.~\cite{AlvarezGaume:1981hn} for sigma model applications): The linear term in definition~\eqref{defBeta} is chosen such that the $\ge^1$ terms cancels out. The $\ge^0$ terms show that the single pole counter terms $T_{(1)}(H)$ dictate the form of the beta functions
\equ{
\gb(H)_i = - \Big[ 1 - H\cdot \pp{H} \Big] T_{(1)}(H)_i 
= \sum_{L \geq 1} L\, T_{(1,L)}(H)_i~. 
}
Given that vertices are expressed as derivatives of the functions $H_i$ and propagators are inverses of similar objects, the operator $H\cdot \pp{H}$ has $+1$ eigenvalue on vertices and $-1$ on propagators. In the last equality we used this observation together with the topological identity $\#(\text{propagators})-\#(\text{vertices})=L-1$ holds for an arbitrary $L$ loop graph. The $1/\ge^n$ poles lead to recursion relations,
\equ{
 - \Big[ 1 - H\cdot \pp{H} \Big] T_{(n+1)}(H)_i  = 
 \gb(H) \cdot \pp{H} T_{(n)}(H)_i~, 
 \label{Recursion} 
}
so that the $T_{(n)}(H)$ for $n\geq 2$ can be expressed in terms of the single pole contributions $T_{(1)}(H)$.

\subsubsection*{Finiteness}

To investigate the finiteness of the (2,0) theory it is sufficient to determine the conditions for which the field dependent beta functions $\gb(H_i)$ are trivial~\cite{Friedan:1980jf}. Trivial beta functions do not necessarily vanish: A two dimensional theory is finite when all $1/\ge$ divergences in $\gb(H)$ vanish up to gauge transformations~\cite{Hull:1985zy}. In the (2,0) context this means that we can allow for divergent holomorphic redefinitions~\eqref{redefKbK} and~\eqref{redefMbM}, target space diffeomorphisms~\eqref{diffTarget} and gauge transformations \eqref{gauge}. The resulting conditions are sometimes called vanishing of the ``B--functions'', which have been derived in components and (1,1) superspace in e.g.~\cite{Hull:1985rc,Hull:1985zy,Hull:1986hn}. For our (2,0) theories we obtain the finiteness conditions 
\begin{subequations}
\label{Finiteness} 
\equa{ 
\gb(K)_a &= 2\, G_{\ub a}\, \bff^\ub_{(1)}  + 2\, B_{ab}\, f^b_{(1)} + k_{(1)a}~, 
\label{FinitenessK} 
\\[2ex]
\gb(N)\phantom{{}_a} &= \bg_{(1)}\, N + N\, g_{(1)} + 
N_{,a}\, f^a_{(1)} + N_{,\ua}\, \bff^\ua_{(1)}~,  
\label{FinitenessN} 
\\[2ex]
\gb(M)\phantom{{}_a} &= g^T_{(1)}\, M + M\, g_{(1)} + 
M_{,a}\, f^a_{(1)} + M_{,\ua}\, \bff^\ua_{(1)} + m_{(1)}~.
\label{FinitenessM} 
}
\end{subequations} 
Here $f_{(1)}$ and  $g_{(1)}$ denote the possible holomorphic single pole renormalizations 
\equ{ 
f(\gf) = \gf + \frac 1\ge\, f_{(1)}(\gf) + \cO\big( \frac 1{\ge^2}\big)~. 
\qquad 
g(\gf) = \Id + \frac 1\ge\, g_{(1)}(\gf) + \cO\big( \frac 1{\ge^2}\big)~. 
}
in holomorphic diffeomorphisms~\eqref{diffTarget} and gauge transformations~\eqref{gauge}.  In addition the beta functions for $K$ and $M$ are only defined up to holomorphic reparameterizations~\eqref{redefKbK} and~\eqref{redefMbM}, 
\equ{
k(\gf) = \frac 1\ge\, k_{(1)}(\gf) + \cO\big( \frac 1{\ge^2}\big)~, 
\qquad 
m(\gf) = \frac 1\ge\, m_{(1)}(\gf) + \cO\big( \frac 1{\ge^2}\big)~, 
}
hence they may lead to the renormalizations $k_{(1)}$ and $m_{(1)}$ given in~\eqref{FinitenessK} and~\eqref{FinitenessM}, respectively.

As emphasized in the references above, contrary to the requirement of vanishing beta functions, the finiteness conditions~\eqref{Finiteness} are off--shell and regularization scheme independent. The equations stated in~\eqref{Finiteness} are superspace equations, and as such still encode crucial requirements for when finiteness is compatible with (2,0) off--shell worldsheet supersymmetry.\footnote{The importance of such superspace conditions was recently re--emphasized in the related (1,1), (2,1) and (2,2) superspace context in~\cite{Hull:2012dy}.} Because of the arbitrary renormalization functions $f_{(1)}$, $g_{(1)}, k_{(1)}$ and $m_{(1)}$ the finiteness conditions~\eqref{Finiteness} might not seem very stringent requirements. However, because these functions all have to be  holomorphic, while the beta functions, the metric, $B_2$--field and the functions $N$ and $M$ are all non--holomorphic, the finiteness conditions~\eqref{Finiteness} can essentially only be solved for $f_{(1)}=g_{(1)}=0$.

\subsubsection*{Quantum superfield tadpoles} 

Not only the action for the background fields renormalizes, also the quantum superfield action does. However, essentially all renormalizations of the quantum action can be ignored for the purposes of this work, with one important exception: tadpoles for the quantum superfields $\gF$ and $\bgF$. 
The quantum superfield tadpoles 
\equ{
S_\text{tadp} = \int\d^D\d^2\gth^+\,
\frac i4 \Big( \bY_{(0)\ua}\, \derr \bgF^\ua - Y_{(0)a}\, \derr \gF^a \Big)~, 
\qquad 
Y_{(0)} =  \gm^{2\ge}\, \sum_{n\geq 1} \frac 1{\ge^n}\, T_{(n)}(Y)~,
\label{QuantumTadpoles} 
}
for $\gF$ and $\bgF$ might be generated at the loop level. Since these tadpoles are absence at tree level, there is no renormalized $Y$ and $\bY$. The dependence of the bare function $Y_{(0)}$ of the renormalization scale gives
\equ{
\ge\, Y_{(0)} + \gb(H) \cdot \pp[Y_{(0)}]{H} = 0~, 
}
which gives a similar recursion as~\eqref{Recursion} for $T_{(n)}(Y)$, $n \geq 2$. Finiteness requires that the single pole contribution $T_{(1)}(Y)$ is holomorphic.

\subsection{Weyl anomaly} 
\label{sc:SuperWeylAnom}

\subsubsection*{Combinded Weyl invariance and (2,0) finiteness}

We investigate the consequence of combining the (2,0) finiteness conditions derived in the previous subsection with the requirement of invariance under the standard Weyl transformations~\eqref{Weyl}. (This repeats the analysis of~\cite{Hull:1985rc} but uses the consequence of (2,0) supersymmetry of the matter sector.)

In $D=2-2\ge$ dimensions the worldsheet theory is not invariant under Weyl transformation~\eqref{Weyl} anymore, this constitutes the Weyl anomaly. For the terms of the bare action involving the Fermi multiplet this anomaly takes the form 
\equ{
\int \d^2\gs\, (-2\ge\, \go) \int\d^2\gth^+\, \frac 12 
\Big\{ \bgL\, N_0\, \gL +\sfrac{1}{2}\, \gL^T \, M_0\,\gL
+\sfrac{1}{2}\, \bgL\, \overline{M}_0 \,\bgL{}^T\Big\}~. 
}
Next we use~\eqref{NoScaleH0} in the form  
\equ{
- \ge\, H_{(0)i} = \gb(H) \cdot \pp[H_{(0)i}]{H} = 
\gm^{2\ge} \gb(H)_j \Big( \gd_{ji} + 
\sum_{n\geq1} \frac 1{\ge^n}\, \pp{H_j} T_{(n)}(H)_i 
\Big)~
\label{WeylAnomalySource}
}
and replace the beta functions $\gb(N)$ and $\gb(M)$ by their structure required by finiteness, i.e.\ equations~\eqref{FinitenessN} and~\eqref{FinitenessM}. If we then use that the Weyl parameter $\go(\gs)$ is an arbitrary function on the worldsheet, implies that the holomorphic functions $f_{(1)}(\gf)=g_{(1)}(\gf)=m_{(1)}(\gf)=0$. Hence we conclude that the beta functions for $N$ and $M$ vanish identically: 
\equ{ 
\gb(N)(\gf,\bgf) = 0~, 
\qquad 
\gb(M)(\gf,\bgf) = 0~. 
\label{BetakNM} 
} 

For the chiral multiplet consequences of Weyl invariance is slightly more complicated because the target space coordinates also appear in the dilaton $\gPs$. Following the same procedure as above for the Fermi multiplets we obtain from~\eqref{FinitenessK} 
\equ{
\int \d^2\gs\,  \Big\{
2 \go \int\d^2\gth^+\, 
 \Big( 
\frac i4\, \der_R \bgf^\ua\, \bk_{(1)\ua}(\bgf)
- \frac i4\, \der_R \gf^a\, k_{(1)a}(\gf)  
\Big) 
+ \frac 1{2\gp} \, \der_L \der_R \go\, \gPs(z, \bz)
\Big\} =0~, 
}
where we only kept the finite terms in $1/\ge$ for simplicity. The last term results form the Weyl transformation of the Einstein--Hilbert term from a flat worldsheet given in~\eqref{Weyl}. If we then evaluate the superspace integrals using that the integrand are purely (anti--)holomorphic, we find that they just result in a $\der_L$ derivative. Normally one does not care about this derivative, because it just gives a total derivative under the worldsheet space--time integral. However, because of the local Weyl parameter $\go(\gs)$ this argument does not apply here. Instead if we perform partial integrations such that in all term there is a $\der_L$ on the Weyl parameter, we find: 
\equ{
\int \d^2\gs\, \der_L\go\, \Big\{
\int\d^2\gth^+\, 
\frac 12\, \der_R z^a\, k_{(1)a}(z)
+ \frac 12\, \der_R \bz^\ua\, \bk_{(1)\ua}(\bz)  
-\frac1{2\gp}\, \der_R z^a\,  \gPs_{,a} -\frac1{2\gp}\, \der_R\bz^\ua\, \gPs_{,\ua}
\Big\} =0~. 
}
We thus conclude that the renormalization one--forms $k_{(1)}$ and $\bk_{(1)}$ are not zero, but determined as the (anti--)holomorphic derivatives of the dilaton: $\gp\, k_{(1)a} = \gPs_{,a}$ and $\gp\, \bk_{(1)\ua} =\gPs_{,\ua}$. Hence we find the conditions
\equ{
\gp\, \gb(K)_a(z,\bz) - \gPs_{,a}(z,\bz) = 0~, 
\qquad
\gp\, \gb(\bK)_\ua(z,\bz) - \gPs_{,\ua}(z,\bz) = 0~, 
\label{BetaK}
\qquad \gPs_{,\ua a}(z,\bz) = 0~. 
} 
The last equation follows because the one--forms $k_{(1)}$ (and $\bk_{(1)}$) are (anti--)holomorphic.

Absence of Weyl anomalies should not only be enforced on the background operators, but also on the quantum operators. In particular, the loop--generated quantum superfield tadpoles~\eqref{QuantumTadpoles} lead to additional contributions to the Weyl anomaly, which can be cancelled by expanding the Weyl--varied dilaton action~\eqref{DilatonAction} to first order in the quantum fields as well. By the same arguments as above, this lead to the conditions 
\equ{
\gp\, T_{(1)}(Y)_a - \gPs_{,a} = 0~, 
\qquad 
\gp\, T_{(1)}(\bY)_\ua - \gPs_{,\ua} = 0~. 
\label{WeylTadp}
}

The Weyl invariance conditions~\eqref{BetakNM},~\eqref{BetaK} and~\eqref{WeylTadp} are stronger than the finiteness conditions obtained in the previous Subsection. In particular, we see that indeed $f_{(1)}=g_{(1)}=m_{(1)}=0$ in~\eqref{Finiteness}, as was suggested by holomorphicity considerations in the previous Subsection. In addition, we infer that $k_{(1)_a} \sim \gPs_{,a}$. This is compatible with the holomorphicity of $k_{(1)}$ because of the last condition in~\eqref{BetaK}. Hence, Weyl invariance selects a very specific set of (2,0) finiteness conditions~\cite{Hull:1985rc}.

\subsubsection*{Super Weyl anomaly}

Next we show that the conditions found above can be very elegantly be derived by using (2,0) super Weyl invariance. Because the supergravity measure $\cE$ is the determinant of the super vielbein, its super Weyl transformation~\eqref{superWeyl} is modified to 
\equ{
 \gd_\cS \cE =
 (1-\ge) \cS\, \cE~ 
\label{MeasureDtrans}
}
in $D=2-2\ge$ dimensions. Hence, if we perform a super Weyl transformation from a locally flat worldsheet $\cG_R = 0$, we not only get contributions from the Einstein--Hilbert term~\eqref{DilatonAction}, but also from the full bare action~\eqref{ClAction}. These additional contributions proportional to $\ge$ constitute the super Weyl anomaly.

Keeping only the finite contributions in the limit $\ge\ra 0$, we obtain the condition for the background
\equa{
\int\d^D\gs\d^2\gth^+\, \Big\{&
\frac{1}{2}\, \cS\, \big( 
\bgL\, \gb(N)\, \gL +\sfrac{1}{2}\, \gL^T \, \gb(M)\,\gL
+\sfrac{1}{2}\, \bgL\, \gb(\overline{M}) \,\bgL{}^T 
\big) + 
\non \\[2ex] 
& - \frac{i}{4}\, \cS\, \big( 
\derr\bgf^\ua\, \gb(\bK)_\ua - \gb(K)_a\,\derr\gf^a 
\big) 
- \frac 1{4\gp}\, \cU\, \big( 
\gPs_{,a}\, \derr\gf^a + \gPs_{,\ua}\, \derr\bgf^\ua
\big)
\Big\} = 0~,
\label{CompsuperWeyl}
}
Since $\gL, \bgL$ are arbitrary Fermi superfields and $\cS$ an arbitrary super Weyl parameter, this leads to the superfield conditions~\eqref{BetakNM}. For the pure chiral superfield contributions the analysis is slightly more complicated because both $\cS$ and $\cU$ appear. However, using that  their super covariant derivatives are related~\eqref{SUconstr}, one finds by picking out their fermionic components the conditions~\eqref{BetaK}.

\subsection{One loop renormalization}
\label{sc:Renorm1L}

\begin{figure}
\begin{center} 
\tabu{ccccccc}{
 \begin{fmfgraph}(12,20) 
 \fmfleft{i} 
 \fmfright{o} 
 \fmf{fermion}{i,o} 
 \fmf{scalar,tension=.7,right=1}{o,o}
  \fmfv{decor.shape=cross,decor.size=4thick}{i,o}
\end{fmfgraph}
&\qquad\qquad&
 \begin{fmfgraph}(12,20) 
 \fmfleft{i} 
 \fmfright{o} 
 \fmf{fermion}{o,i} 
 \fmf{scalar,tension=.7,right=1}{o,o}
  \fmfv{decor.shape=cross,decor.size=4thick}{i,o}
\end{fmfgraph}
&\qquad\qquad\qquad&
 \begin{fmfgraph}(12,20) 
 \fmfleft{i} 
 \fmfright{o} 
 \fmf{fermion}{i,o} 
 \fmf{scalar,tension=.7,right=1}{o,o}
  \fmfv{decor.shape=cross,decor.size=4thick}{o}
\end{fmfgraph}
&\qquad\qquad&
 \begin{fmfgraph}(12,20) 
 \fmfleft{i} 
 \fmfright{o} 
 \fmf{fermion}{o,i} 
 \fmf{scalar,tension=.7,right=1}{o,o}
  \fmfv{decor.shape=cross,decor.size=4thick}{o}
\end{fmfgraph}
\\[-3ex] 
}
\end{center}
\caption{These diagrams describe the one loop corrections to the chiral superfield action. The former two define a tadpole for the background while the latter two for the quantum superfields.
\label{fg:1LChiralRenorm}}
\end{figure}
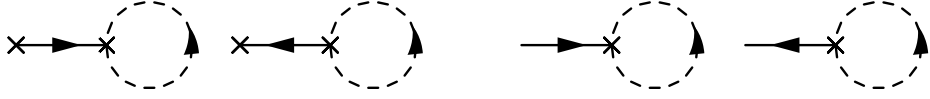

\subsubsection*{Background tadpole}

The one loop corrections to the pure chiral superfield action $S_\gf$ are induced by the first two diagrams displayed in Figure~\ref{fg:1LChiralRenorm}. Using the super propagator~\eqref{ChiralProp} and the two--point insertions~\eqref{ChiralVertices} these diagrams are readily computed to give 
\equ{ 
i\, \gG_{\gf}^{1L} = 
\frac i4\, \Big[ 
\bgG_{\!+}{}^\ua_{\ub\,\ua} \, \derr\bgf^\ub 
- \gG_{\!+}{}^a_{ba} \, \derr\gf^b 
\Big]_2\, 
\gd_{12}^D\, \Big[\frac 1{-\Box_{D} + m^2}\Big]_1 \gd_{12}^D~. 
} 
Consequently their one loop counter terms read 
\equ{ 
\gD_{(1,1)} S_{\gf}^{1L} = - \frac i4
\int \d^2\gs \d^2 \gth^+\, 
\frac {\gm^\ge}{4\gp} \Big[\frac 1\ge\,  + \ln \frac{\bgm^2}{m^2} \Big]
\Big( 
\bgG_{\!+}{}^\ua_{\ub\,\ua} \, \derr\bgf^\ub
- \gG_{\!+}{}^a_{ba} \, \derr\gf^b 
\Big)~, 
\label{BackgroundTadpole1L} 
}
using~\eqref{Simple1L}.

\subsubsection*{Quantum superfield tadpole}

As discussed in Subsection~\ref{sc:EffectiveAction} it is not possible to define a normal coordinate superfield w.r.t.\ to the torsion connection as this would be incompatible with (2,0) supersymmetry. But by using the holomorphic normal coordinates~\eqref{HolomorphicNormalCoordinates} we were able to find a classical/quantum splitting which is manifestly coordinate covariant. As a consequence not all cubic vertices are absent; those involving the torsion remain, see~\eqref{ChiralVertices}. As the latter two diagrams in Figure~\ref{fg:1LChiralRenorm} show, these torsion vertices induce tadpoles for the quantum superfields at one loop. 
\equ{
i\, \gG_{\gF}^{1L} = 
\frac i8\, \Big[ 
  \bH_{\ub\,\ua a}\, G^{a\ua}\, \derr\bgF^\ub 
  - H_{ba\ua}\, G^{a\ua} \, \derr\gF^b 
\Big]_2\, 
\gd_{12}^D\, \Big[\frac 1{-\Box_{D} + m^2}\Big]_1 \gd_{12}^D~. 
} 
This requires the counter terms to take the form: 
\equ{ 
\gD_{(1,1)} S_{\gf}^{1L} = - \frac i8
\int \d^2\gs \d^2 \gth^+\, 
\frac {\gm^\ge}{4\gp} \Big[\frac 1\ge\,  + \ln \frac{\bgm^2}{m^2} \Big]
\Big( 
\bH_{\ub}\, \derr\bgF^\ub  - H_{b} \, \derr\gF^b 
\Big)~, 
\label{QuantTadp1L} 
}
where $H_b = H_{ba\ua}\, G^{a\ua}$, $\bH_\ub =  \bH_{\ub\,\ua a}\, G^{a\ua}$. The presence of tadpole for quantum superfields signifies that the 1PI graphs also include graphs which can become disconnected with such a tadpole split off.

\subsubsection*{Fermi multiplet renormalization}

\begin{figure}[t] 
\begin{center} 
\tabu{ccc}{
 \begin{fmfgraph}(20,20)
 \fmfleft{i} 
 \fmfright{o} 
 \fmf{fermion}{i,v1} 
 \fmf{fermion}{v1,o}
 \fmfdot{v1} 
\fmf{scalar,tension=.6}{v1,v1} 
 \end{fmfgraph}
 \qquad & \qquad 
  \begin{fmfgraph}(30,30) 
   \fmfleft{i} 
 \fmfright{o} 
 \fmflabel{$e^-$}{i}
  \fmf{fermion,tension=2}{i,v1}
 \fmf{fermion,right=.7}{v1,v2} 
 \fmfdot{v1,v2} 
 \fmf{fermion,tension=2}{v2,o}
\fmf{scalar,right=.7,tension=.1}{v2,v1} 
 \end{fmfgraph}
 \qquad & \qquad   
 \begin{fmfgraph}(30,30)
  \fmfleft{i} 
 \fmfright{o} 
 \fmf{fermion,tension=2}{i,v1}
 \fmf{fermion,left=.7}{v2,v1} 
 \fmfdot{v1,v2} 
 \fmf{fermion,tension=2}{v2,o}
\fmf{scalar,left=.7,tension=.1}{v1,v2} 
 \end{fmfgraph}

 \\[-3ex] 
 a & b & c 
  }
\end{center}
\caption{These Figures display the three one--loop diagrams that contribute to the renormalization of the $\bgL\gL$--part of the Fermi superfield action. 
\label{fg:1LFermiRenorm}}
\end{figure}
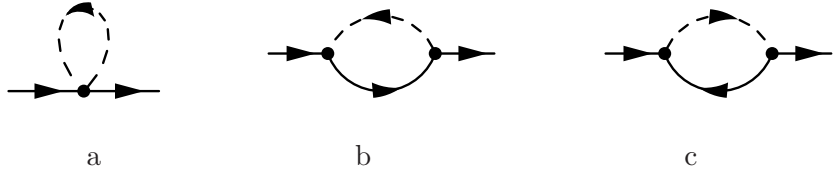

At the one loop level the $\bgL\gL$--part of the Fermi superfield action gets corrected by quantum corrections encoded in Figure~\ref{fg:1LFermiRenorm}. The corresponding propagators and vertices can be found in Subsections~\ref{sc:Propagators} and~\ref{sc:Vertices}, respectively. The first diagram, Figure~\ref{fg:1LFermiRenorm}.a, is a tadpole graph; there is no momentum flowing in or out of the loop. Hence this one loop diagram is readily computed using the dimensional reduction procedure outlined in Appendix \ref{sc:DimRed}. The other two diagrams, Figures~\ref{fg:1LFermiRenorm}.b and c are proper self--energy supergraphs. However, using the result of Subsection~\ref{sc:Contraction} that Fermi superfield lines can be contracted.  Consequently, after some standard supergraph manipulations the first three supergraphs in Figure~\ref{fg:1LFermiRenorm} reduce to the same scalar graph up to pre--factors 
\equ{
i\, \gG_{\bgL\gL}^{1L} = 
\frac 12\, 
\Big[
\bgL{}^\uga \gL^\ga\, G^{a \ua} \Big\{
N_{\uga\ga,\ua a} - N_{\uga\gg, \ua} N^{\gg\ugg} N_{\ugg\ga,a} 
+ \bM_{\uga\ugg, a} N^{\gg\ugg} M_{\ga\gg,\ua} 
\Big\}  
\Big]_2\, \gd_{12}^D\, \Big[\frac 1{-\Box_{D} + m^2}\Big]_1 \gd_{12}^D~,  
}
Note that the expression between the curly brackets $\{\ldots\}$ precisely equals the field strength component $[F_{\ua a}(N)]_{\uga\ga}$ see \eqref{GaugeFieldStrength}.

The super Feynman diagrams that renormalize the $\gL\gL$--part of the kinetic action of the Fermi superfield action are essentially the same as those displayed in Figure~\ref{fg:1LFermiRenorm} but with the external lines both pointing outwards. Hence via similar computations as above
the sum of these three graphs gives 
\equ{
i\, \gG_{\gL\gL}^{1L} = 
\frac 14\, 
\Big[
\gL{}^\ga \gL^\gb\, G^{a \ua} \Big\{
M_{\ga\gb,\ua a} - M_{\ga\gg, \ua} N^{\gg\ugg} N_{\ugg\gb,a} 
+ N_{\ugg\ga, a} N^{\gg\ugg} M_{\gg\gb,\ua} 
\Big\}  
\Big]_2\, \gd_{12}^D\, \Big[\frac 1{-\Box_{D} + m^2}\Big]_1 \gd_{12}^D~,  
}
In this case we can recognize the anti--symmetric tensor part of the target space gauge field strength $[F_{\ua a}(M)]_{\ga\gb}$ given in~\eqref{GaugeFieldStrength}.

After regularizing these diagrams using dimensional reduction and including the tadpole graphs, we determine the counter terms that cancel the divergences at one loop: 

\equ{ 
\gD_{(1,1)} S_{\gL}^{1L} = 
\frac {1}2
\int \d^2\gs \d^2 \gth^+\, 
\frac{\gm^\ge}{4\gp} \, \Big[\frac 1\ge\!+\! \ln \frac{\bgm^2}{m^2}\Big]\, 
G^{a \ua} \Big\{ \bgL\, F_{\ua a}(N)\, \gL 
+\frac{1}{2}\, \gL^T \, F_{\ua a}(M) \,\gL
+\frac{1}{2}\, \bgL\, F_{\ua a}(\overline{M})\,\bgL{}^T \Big\}~. 
\label{FermiRenorm1L} 
}

\subsection{Weyl invariance at one loop}

We consider the requirements of Weyl invariance at the one loop level. Using the results of the previous subsection, we can explicitly determine these conditions. Given that these equations have interesting target space interpretations, we would like to discuss them in turn.

From the one loop renormalization of the Fermi multiplets~\eqref{FermiRenorm1L} we obtain the requirements
\equ{
G^{a\ua}\,F_{\ua a}(N) = 0~, 
\qquad 
G^{a \ua}\, F_{\ua a}(M)  = 0~, 
\label{BetaNM1L} 
}
using the field strengths $F_{\ua a}$ are given in~\eqref{GaugeFieldStrength}. They constitute the first order Hermitean--Yang--Mills equations, rather than the second order Maxwell equations which is obtained by the conventional beta function approach. In target space they are obtained by requiring that the gaugino supersymmetry variation vanishes.

Similar, from the one loop quantum tadpole~\eqref{QuantTadp1L} combined with expanding the  dilaton action~\eqref{DilatonAction} to first order in the quantum superfields, we obtain
\equ{
\frac 1{8}\,  H_a + \gPs_{,a} =0~, 
\label{BetaKQuantTadp1L} 
}
where $H_a = G^{b\ub} H_{ab\ub}$ is the traced torsion tensor. This is precisely the condition one obtains by demanding that the supersymmetry variation of the dilatino vanishes. This equation in particular says that if the target space possesses torsion the dilaton cannot be constant.

Finally, the one loop tadpole~\eqref{BackgroundTadpole1L} for the background fields imply that 
\equ{ 
\frac 1{4}\, \gG_{\!+}{}_{a} + \gPs_{,a} = 0~, 
\label{BetaKTadpole1L} 
} 
with the traced connection $\gG_{\!+}{}_a = (\ln\det G)_{,a} + H_a$, see Appendix~\ref{sc:Connections}. If the target space does not possess torsion this equation and its conjugated imply 
$(\ln\det G)_{,a} = (\ln\det G)_{,\ua} = 0$
using~\eqref{TracedConnections}. Hence the geometry is Ricci flat $R_{\ua a} = (\ln\det G)_{,\ua a} =0$ 
and hence Calabi--Yau. If torsion is present, we can combine the conditions in~\eqref{BetaKTadpole1L} and~\eqref{BetaKQuantTadp1L} to find that
\equ{
e^{-4\, \gPs}\, \det G = \text{const}~. 
}
This is a condition that arises for conformally balanced geometries, as a consequence of the gravitino and dilatino supersymmetry variations~\cite{Strominger:1986uh,Becker:2007zj,Becker:2009df}.

Hence observe that in our approach we obtain the conditions that the supersymmetry variations of the target space gauginos and dilatino vanish directly. (That the gauge bundle is holomorphic is built in from the very beginning by demanding (2,0) supersymmetry.) On the other hand we do not obtain the gravitino variation itself, but only some of its consequences. The reason why we do not find the gravitino variation itself is that it involves a space--time derivative on the local supersymmetry parameter. In the RNS--formulation, which we have employed throughout this paper, target space supersymmetry is not made manifest, hence this formulation does not contain an obvious representative of this parameter. One of the consequences of the gravitino variation, namely that the geometry is complex, we have used as a central input: We have enforced that the worldsheet possesses (2,0) supersymmetry which immediately requires that the target space is complex. However, the other consequences of the gravitino variation are encoded in~\eqref{BetaKTadpole1L}.

This analysis establishes that by demanding (super) Weyl invariance for a (2,0) worldsheet gives rise to the consequence BPS equations directly. This should be put in contrast with the standard way the target supersymmetry conditions are obtained from the heterotic string: First the target space effective action has to be determined. This can either be done by lifting the target space equations of motions, which are obtained by the requirement of vanishing beta functions, to an effective action, or by computing string scattering amplitudes. Next, one determines which target space on--shell supersymmetry variations are compatible with this effective action. By putting the supersymmetry variations of the fermion fields to zero finally gives the BPS equations.

\section{Conclusions}
\label{sc:Concl}

We have considered the heterotic string in the RNS--formulation. We have assumed that the corresponding NLSM possesses (2,0) supersymmetry, so that it describes a complex torsion target space that supports a holomorphic vector bundle. Both the metric and $B_2$--field are determined by a one--form pre--potential $K_a$ and its conjugate $\bK_\ua$. The gauge background is encoded by matrix--valued pre--potentials $N$, $M$ and $\bM$. The classical (2,0) NLSM is completely specified by these pre--potentials.

This worldsheet theory was quantized using the path integral. By utilizing holomorphic normal coordinates we ensured that the resulting effective action gives rise to covariant supergraphs. To regularize the theory we made use of a variant of dimensional regularization so that the ultra--violet divergences show up as $1/\ge$ poles. This scheme allows for a systematic renormalization of the theory at the loop level, which enables us to determine the conditions that the beta functions associated to these pre--potentials are trivial. Even though in principle trivial not necessarily means zero, holomorphicity of the target space diffeomorphisms and gauge transformations, implies that the beta function for $N$ is identically zero, while those of $K_a$ and $M$ vanish up to possible holomorphic contributions.

Our main result states that the cancellation of the Weyl anomaly combined with the conditions for finiteness of the (2,0) NLSM implies the heterotic BPS conditions, i.e.\ the conditions for unbroken supersymmetry in four dimensions. An elegant way to derive these conditions is to enforce (2,0) super Weyl invariance. In our approach we obtain the conditions that the supersymmetry variations of the target space gauginos and dilatino vanish directly. In particular the HYM equations are simply the superfield beta functions set to zero. On the other hand we do not obtain the gravitino variation itself. 
One of its consequences, the fact that the geometry has to be complex, was used to motivate that the worldsheet possesses (2,0) supersymmetry.  
However, the remaining consequences of the gravitino variation are recovered. At the one loop level we find that the geometry is required to be conformally balanced, and without torsion the background becomes Calabi--Yau.

The explicit one loop renormalization of a (2,0) NLSM can be extended to higher orders. This means that one can directly compute $\alpha'$ corrections to the supersymmetry conditions. (We plan to present two loop results and discuss anomalies in a subsequent publication.) 
Moreover, it might even be possible to extend our results non--perturbatively along the lines of the (2,0) non--renormalization theorem  recently derived in~\cite{Cui:2011rz,Cui:2011uw}. 
This should be contrasted with the standard procedure, in which first the effective action has to be determined via either string scattering amplitudes or beta function computations to a given order in $\alpha'$, and after that determine the associated supersymmetry variations that leave this effective action invariant. This is particular difficult as the ten dimensional supersymmetry is only on--shell, hence corrections to the effective action can potentially induce modifications to the supersymmetry variations. Our approach completely by passes all these complications and gives the BPS conditions directly.

Our results might shed some new light on the question what the requirement of having a stable holomorphic vector bundle means from the worldsheet perspective. Stability is the mathematical condition that ensures that the HYM equations can be solved on a compact CY manifold. Hence from the worldsheet the relevant condition seems to be that the theory is finite in a (2,0) supersymmetric fashion and possesses Weyl invariance at the quantum level. This is a condition that can be tested even in regions of the moduli space where string corrections to the geometry become large.

\subsection*{Acknowledgments}

We would like to thank David Andriot, Michael Kay, Peter Patalong, Daniel Plencner and Thorsten Rahn for stimulating discussions. 
Especially we would like to thank Chris Hull for very helpful discussions at the initial stage of this project and pointing out ref.~\cite{Hull:1986hn} to us. 
Moreover, we would like to thank Ilarion Melnikov and Savdeep Sethi for valuable comments to the first version of this manuscript. 
This work has been supported by the LMUExcellent Programme.

\appendix 
\def\theequation{\thesection.\arabic{equation}} 

\setcounter{equation}{0}

\section{(2,0) Superspace conventions}
\label{sc:Conventions}

Our superspace conventions are compatible with Wess and Bagger \cite{wundb} unless stated otherwise. For completeness we summarize them here.

The Minkowski metric on the worldsheet with coordinates $\gs =(\gs^0,\gs^3)$ has the signature $(-,+)$. We introduce light--cone coordinates on the worldsheet 
$\gs^L = \half(\gs^0+\gs^3)$ and $\gs^R = \half(\gs^0-\gs^3)$ with the corresponding left-- and right--moving derivatives 
\equ{ 
\derr = \der_0 - \der_3~, 
\qquad 
\derl = \der_0 + \der_3~, 
\qquad 
\Box_2 = - \derl\derr = -\der_0^2 + \der_3^2~, 
}
with $\der_0 = \frac{\der}{\der \gs^0}$ and $\der_3 = \frac{\der}{\der \gs^3}$, such that $\derl\gs^L=\derr\gs^R=1$ and $\derl\gs^R = \derr\gs^L=0$.

\subsubsection*{Super covariant derivatives}

 The worldsheet coordinates $\gs$ are amended by Grassmann variables $\gth = (\gth^+,\bgth^+)$ to form superspace coordinates $(\gs, \gth) = (\gs^0,\gs^3,\gth^+,\bgth^+)$. The supercovariant derivatives, 
\equ{ 
D_+ = \frac{\der}{\der \gth^+}-i\bgth^+\derl~, 
\qquad 
\bD_+ = -\frac{\der}{\der \bgth^+}+i\gth^+\derl~, 
\label{superD}
}
satisfy the following D--algebra properties:
\equ{ 
\arry{c}{\dsp 
\{D_+,\bD_+\} = 2i \derl~, 
\qquad 
D_+^2 = \bD_+^2 = 0~, 
\\[2ex] \dsp 
D_+ \bD_+ D_+ = 2i \derl\, D_+~, 
\qquad 
\bD_+ D_+ \bD_+ = 2i \derl\, \bD_+~. 
}
\label{AlgebraD} 
}
Denote by $C$ and $\bC$, a chiral superfield ($\bD_+ C = 0$), and an anti--chiral superfield ($D_+ \bC =0$), respectively. Since any (anti--)chiral superfield can be expressed as 
$C = \bD_+ F$ ($\bC = D_+ F$) in terms of an unconstraint superfield $F$, the latter two D--algebra properties imply that 
\equ{ 
\bD_+D_+\, C = 2i\derl\, C~,
\qquad 
D_+\bD_+ \, \bC = 2i\derl\,\bC~. 
\label{bDDonChiral}
}

\subsubsection*{Superspace integration} 

The full superspace integration over an arbitrary superfield $F$ can be rewritten as 
\equ{
\int \d^2 \gs \d^2\gth^+\, F = \int\d^2\gs\, \frac 12\, [D_+,\bD_+] F(\gs,\gth)|~. 
\label{FullInt} 
}
Here after Grassmannian differentiation the Grassmann variables $\gth^+, \bgth^+$ are set to zero; this is denoted by $|$. In addition to the full superspace integration we define a (anti--)chiral superspace integral over any (anti--)chiral superfield 
\equ{
\int \d^2 \gs \d\gth^+\, C = \int\d^2\gs\, D_+ C(\gs,\gth)|~, 
\qquad 
\int \d^2 \gs \d\bgth^+\, \bC = -\int\d^2\gs\, \bD_+ \bC(\gs,\gth)|~, 
\label{ChiralInt} 
}

\subsubsection*{Superspace delta function}

For two coordinate systems $(\gs,\gth)_1$ and $(\gs,\gth)_2$ we define the superspace delta function 
\equ{ 
\gd_{12} = 
(\gth^{+}_1-\gth^{+}_{2})(\bgth^{+}_{1}-\bgth^{+}_{2})\, \gd^2_{12}~, 
\qquad 
\gd^2_{12} = \gd^2(\gs_1-\gs_2)
\label{superDelta} 
}
hence $\gd^2_{12}$ denotes the usual two dimensional delta function on the worldsheet. Here and throughout the paper we  employ the short--hand notation using labels to indicate the coordinate systems in which the corresponding objects are evaluated, e.g.\ $F_1 = F(\gs_1,\gth_1)$. The ordering of the $(\gth^+_1-\gth^+_2)$ and $(\bgth^+_1-\bgth^+_2)$ factors in~\eqref{superDelta} has been chosen to be compatible with the ordering of the Grassmannian integration in \eqref{FullInt}, i.e.\ such that the defining property of the delta function 
\equ{ 
\int [\d^2 \gs \d^2\gth^+]_1 \, \gd_{12}\, F_{1} = F_{2}~,  
}
holds without any additional signs.

One can change the coordinate system of a (super covariant) derivative which acts on a delta function at the cost of a sign: 
\equ{ 
\arry{c}{ \dsp 
\derl_1\, \gd_{12} = - \derl_2\, \gd_{12}~, 
\qquad 
\derr_1\, \gd_{12} = - \derr_2\, \gd_{12}~, 
\\[2ex] \dsp 
D_{+1}\, \gd_{12} = - D_{+2}\, \gd_{12}~, 
\qquad 
\bD_{+1}\, \gd_{12} = - \bD_{+2}\, \gd_{12}~. 
}
}
When two consequent super covariant derivatives fall on a delta function, changing their coordinate systems is only possible provided that their order is interchanged: 
\equ{ 
\bD_{+1} D_{+1} \,\gd_{12} = - \bD_{+1} D_{+2} \,\gd_{12} =  
D_{+2} \bD_{+1}\, \gd_{12} = - D_{+2} \bD_{+2}\, \gd_{12}
\label{ChangeOfCoords}
}
This again comes at the price of a sign, as they are both fermionic.

An important non--renormalization theorem~\cite{Gates:1983nr} is that any supergraph becomes local in the Grassmann variables. This result can be summarized as follows:
\equa{
- &\int [\d^2 \gs \d^2\gth^+]_{12}\, 
F_2(\gth^+_2,\der_2)\,  \gd_{12}\,  F_1(\gth^+_1,\der_1) [\bD_+D_+]_1\gd_{12} 
= 
\\[1ex] 
&\int\! [\d^2 \gs \d^2\gth^+]_{12}\, 
F_2(\gth^+_2,\der_2)\,  \gd_{12}\,  F_1(\gth^+_1,\der_1) [D_+\bD_+]_1\gd_{12} 
= 
\int\! [\d^2\gs]_{12}\d^2\gth^+\, 
F_2(\gth^+,\der_2)\, \gd^2_{12}\, F_1(\gth^+,\der_1)\,\gd^2_{12}~, 
\non
}
where $F_i(\gth^+_i, \der_i)$ denote generic operators that do not involve any super covariant derivatives.

\setcounter{equation}{0}

\section{Torsion connections}
\label{sc:Connections}

In the main text we have defined various connections. Besides the Hermitean connection $\gG^{b}_{cd}$ two torsion connections 
$\gG_{\!\pm}{}^{b}_{cd}$ were introduced in~\eqref{TorsionConnections}. In general we define 
\equ{
\gG_{\!\ga\,}{}^b_{cd} = G^{b\ub} 
\Big( G_{\ub c,d} + \sfrac{\ga}2\, H_{cd\ub}\Big)~, 
\qquad 
\bgG_{\!\ga\,}{}^\ub_{\uc\,\ud} = G^{b\ub} 
\Big( G_{\uc b,\ud} + \sfrac{\ga}2\, \bH_{\uc\,\ud b} \Big)~, 
} 
for an arbitrary $\ga \in \Real$, i.e.\ so that definitions of $\gG_{\!\pm\,}{}^{b}_{cd}$ agree and the Hermitean connection reads $\gG_{\!0\,}{}^b_{cd} = \gG^{b}_{cd}$. Since the connection are expressed in terms of $H_{ab \ua}$, they invariant under the holomorphic $B$--field transformations~\eqref{holoBtrans}.

Using that both the metric $G$ and the $B$--field are expressed by~\eqref{Metric} and~\eqref{Bfield} in terms of the one--form potentials $K_a$ and $\bK_\ua$, one obtains the following identities 
\begin{subequations} 
\label{ConnectionIds}
\equ{
\gG_{\!\ga\,}{}^b_{cd} - \gG_{\!\gb\,}{}^b_{cd} = \sfrac{\ga -\gb}2\, H_{cd}{}^b~, 
\qquad
\bgG_{\!\ga\,}{}^\ub_{\uc\,\ud} - \bgG_{\!\gb\,}{}^\ub_{\uc\,\ud} = \sfrac{\ga -\gb}2\, \bH_{\uc\,\ud}{}^\ub~, 
\\[2ex]   
\gG_{\!\ga\,}{}^b_{cd} - \gG_{\!\ga\,}{}^b_{dc} = \sfrac{2\ga+1}2\, H_{cd}{}^b~, 
\qquad
\bgG_{\!\ga\,}{}^\ub_{\uc\,\ud} - \bgG_{\!\ga\,}{}^\ub_{\ud\,\uc} = \sfrac{2\ga+1}2\, \bH_{\uc\,\ud}{}^\ub~.
}
\end{subequations}
The identities show that there are two connections in which can be expressed such that the torsion does not explicitly appear:
$\gG_{\!0\,}{}^b_{cd} =  G^{b\ua}\, G_{\ua c, d}$ and
$\gG_{\!-\,}{}^{b}_{cd} = G^{b\ua}\, G_{\ua d, c}$. However, only the connection,  
\equ{
\tgG{}^b_{cd} = \gG_{\!-\sfrac12\,}{}^b_{cd} 
= \frac 12\Big( 
\gG_{\!\ga\,}{}^b_{cd} + \gG_{\!\ga\,}{}^b_{dc}
\Big)~, 
\label{symmConnection}
}
is symmetric in indices $c,d$, hence this connection is neither $\gG_{\!0}$ nor $\gG_{\!-}$.

The connections lead to the following curvatures 
\equ{
R_{\ga\,}{}_{\ua a \ub b} = G_{\ua c} R_{\ga\,}{}^c_{a\ub b}~, 
\quad 
R_{\ga\,}{}^c_{a\ub b} = \gG_{\!\ga\,}{}^c_{ab,\ub}~, 
\qquad 
\bR_{\ga\,}{}_{a \ua b \ub} = G_{\uc a} \bR_{\ga\,}{}^\uc_{\ua b \ub}~, 
\quad 
\bR_{\ga\,}{}^\uc_{\ua b \ub} = \bgG_{\!\ga\,}{}^\uc_{\ua\,\ub,b}~, 
}

From these torsion connections we can define $U(1)$ connections and their curvatures 
\equ{
\gG_{\!\ga\,}{}_a = \gG_{\!\ga\,}{}^b_{ab}~, 
\quad 
\bgG_{\!\ga\,}{}_{\ua} = \bgG_{\!\ga\,}{}^\ub_{\ua\,\ub}~, 
\qquad 
R_{\ga\,}{}_{a\ua} = R_{\ga\,}{}^b_{a\ua b}~,
\quad 
\bR_{\ga\,}{}_{\ua a} = \bR_{\ga\,}{}^\ub_{\ua a \ub}~. 
}
The $U(1)$ connection and curvature with $\ga = -$ can be expressed as derivatives 
\equ{
\gG_{\!-\,}{}_a = (\ln\det G)_{,a}~, 
\quad 
\bgG_{\!-\,}{}_\ua = (\ln\det G)_{,\ua}~, 
\qquad 
R_{-\,}{}_{a\ua} =  \bR_{-\,}{}_{\ua a} = (\ln\det G)_{,\ua a}~,
}
of the determinant of the metric only. These expressions are reminiscent of identities that hold for \Kh\ geometries. The others can be written as 
\equ{
\gG_{\!\ga\,}{}_a = (\ln\det G)_{,a} + \sfrac {\ga+1}2\, H_a~,  
\qquad 
\bgG_{\!\ga\,}{}_\ua = (\ln\det G)_{,\ua} + \sfrac {\ga+1}2\, \bH_\ua~,
\label{TracedConnections} 
}
where $H_a = H_{ab}{}^b$ and $\bH_\ua = \bH_{\ua\,\ub}{}^\ub$. Finally we can define the symmetrized and anti--symmetrized curvatures
\begin{subequations}
\equ{
S_{\ga\,}{}_{\ua a} = 
\bgG_{\!\ga\,}{}_{\ua,a} + \gG_{\!\ga\,}{}_{a,\ua} = 2\,(\ln\det G)_{,\ua a} 
+ \sfrac {\ga+1}2\, \big(\,\bH_{\ua,a} +  H_{a,\ua} \big)~, 
\\[2ex] 
C_{\ga\,}{}_{\ua a} = 
\bgG_{\!\ga\,}{}_{\ua,a} - \gG_{\!\ga\,}{}_{a,\ua} = 
\sfrac {\ga+1}2\, \big(\, \bH_{\ua,a} - H_{a,\ua} \big)~, 
}
\end{subequations}

\setcounter{equation}{0}

\section{Dimensional Regularization}
\label{sc:DimRed}

Regularization by dimensional regularization~\cite{'tHooft:1972fi} is well--known, for a detailed review see e.g.~\cite{Collins:1984xc}. Also in two dimensions this scheme has proven very powerful~\cite{Hull:1987pc,Metsaev:1987zx}. In the dimensional regularization one extends the two dimensional space time integral 
\equ{
\int\d^2\gs \ra
 \int \frac{\d^D\gs}{\gm^{2-D}} = 
 \int \frac{\d^2\gs_\parallel \d^{D-2}\gs_\perp}{\gm^{2-D}}
}
to $D>2$ dimensions and introduce the regularization scale $\gm$ to preserve the mass dimension of the original two dimensional theory. Therefore in $D$ dimensions the Fourier transform  $\tA_1 = A(p_1)$ of $A_1 = A(\gs_1)$ is given by  
\equ{ 
A_1 = \int \frac{\d^D p_1}{(2\gp)^D \gm^{D-2}} \, \tA_1\, 
e^{i p_1\cdot \gs_1}~.  
}
The $D$ dimensional coordinate and momentum delta--functions can be represented as 
\equ{  
\gd^D_{12} = \gd^D(\gs_1-\gs_2) = \int \frac{\d^Dq_1}{(2\gp)^D\gm^{D-2}}\, 
 e^{q_1\cdot(\gs_1- \gs_2)}~, 
\quad 
 \gd^D(p_1- p_2) = \int \frac{\d^D\gs_1}{(2\gp)^D\gm^{2-D}}\,
  e^{\gs_1\cdot(p_1- p_2)}~.  
} 

\subsection*{Clifford algebra in $D$ dimensions}

We recall how fermions are regularized in dimensional regularization \cite{'tHooft:1972fi,Breitenlohner:1977hr,Collins:1984xc}. One defines the gamma matrices $\gG^a$ in $D$ such that 
\equ{
\{ \gG^a, \gG^b \} = 2\, \get^{ab}~. 
}
The chirality operator $\tgG = \gG^0\gG^3$ is defined as in two dimensions which implies that the Lorentz invariance in $D$ dimensions is broken to $SO(1,1)\times SO(D-2)$ since 
\equ{
\{ \tgG, \gG^a \} = 0~, a = 0,3~, 
\qquad 
[ \tgG, \gG^a ] = 0~,~ a = 1,\ldots, D-2~. 
}
We can choose a chiral basis for the gamma matrices in $D$ dimensions by 
\equ{ 
\gG^0 = \gs^1 \otimes \Id~, 
\qquad 
\gG^3 = \gs^2 \otimes \Id~, 
\qquad 
\tgG = \gs^3 \otimes \Id~, 
\qquad 
\gG^a = \gs_3 \otimes \gg^a~,
}
where the gamma matrices $\gg^a$ define a basis for the Clifford algebra in $D-2$ dimensions.

In this basis the action for a Dirac fermion $\gPs^T= (\gl^T, \bgps)$, $\bgPs = -i(\gps^T,\bgl)$ can be decomposed as \cite{wundb}
\equ{ 
S = \int \d^D\gs \Big\{
i\, \bgl \derl \gl + i\, \bgps \derr \gps 
+ \gps^T \big( \der\Slashed_\perp+m \big) \gl 
- \bgl \big( \der\Slashed_\perp -m \big) \bgps^T
\Big\}~.
\label{FermionActionD}
}
This shows that this regularization preserved the notion of left-- and right--moving fermions.

\subsection*{Regularized chiral and Fermi propagators}

To determine the regularized propagators for (2,0) chiral and chiral--Fermi superfields, we supersymmetrize the action \eqref{FermionActionD} describing dimensional regularized left-- and right--moving fermions. Therefore, we consider the theory 
\equ{
S = - \int \d^D\gs \Big\{ 
 \frac 12 \int \d^2 \gth^+
\Big( \bgL \gL + i \bgF \derr \gF \Big)
+ \frac 1{\sqrt 2} \int \d\gth^+\gF\big( \der\Slashed_\perp+m \big)  \gL 
- \frac 1{\sqrt 2} \int \d\bgth^+ \big( \der\Slashed_\perp-m \big) \bgF \bgL
\Big\}~. 
}
Coupling the superfields $\gF$ and $\gL$ to the sources $J$ and $\gO$, respectively, gives the expression 
\equa{
i S_{J,\gO} = & \int\d^2\gs\d^2\gth^+ \, \Big(  
\bJ\, \frac{\derr}{-\Box_D + m^2} \, J 
- i\, \bgO\, \frac 1{-\Box_D + m^2} \, \gO 
\Big) + 
\\[2ex] 
& + \sqrt 2 \int\d^2\gs\d\gth^+\, 
J \frac {\der\Slashed_\perp -m}{-\Box_D + m^2} \gO 
- \sqrt 2 \int\d^2\gs\d\bgth^+\, 
\bJ \frac {\der\Slashed_\perp+m}{-\Box_D+ m^2} \bgO~,
\non 
}
where $\Box_D = - \derl\derr + \der_\perp^2$.

Now we take either $\gF$ or $\gL$ as a physical field and think of the other as regulator field and we only use the source $\gO$ or $J$, respectively. In other words set one of the sources to zero, so that the second line vanishes identically and only one of the two terms in the first line survives. Hence the regularized form of the propagators for the Fermi and chiral superfields are \eqref{FermiProp} and \eqref{ChiralProp}.

\subsection*{Dimensional regularized momentum integrals}

We define the following general $D$ dimensional Euclidean integral
\equa{
I^\gb_\ga(m^2|D)\ &= 
\int \frac{\d^D q}{(2\gp)^D\gm^{D-2}} \frac{(q^2)^\gb}{(q^2 + m^2)^\ga}
=
\int \frac{i\, \d^D q_E}{(2\gp)^D\gm^{D-2}} \frac{(q^2)^\gb}{(q^2 + m^2)^\ga} = 
\non  \\[2ex] 
& =
\frac {i}{4\gp} \, 
\frac{1}{ (m^2)^{\ga - \gb -1}}\,  
\frac{\gG(D/2+\gb)\, \gG(\ga-\gb- D/2)}{\gG(D/2)\, \gG(\ga)} \, 
\Big( 4\gp \frac{\gm^2}{m^2} \Big)^{1-D/2}~.
\label{Dintegrals}
}
The appearance of the ``$i$'' after the second equal sign indicates that we have performed a Wick rotation to Euclidean momentum space. 
In addition we define integrals with integrands proportional $q_\perp^2$: 
\equ{
I^\perp_\ga(m^2|D) = \int \frac{\d^D q}{(2\gp)^D \gm^{D-2}}\, 
\frac{q_\perp^2}{(q^2 +m^2)^\ga} 
= - \frac{i}{4\gp}\,
\frac{1 - D/2}{(m^2)^{\ga-2}} \, 
\frac{\gG(\ga-1-D/2)}{\gG(\ga)}\,  
\Big( 4\gp \frac{\gm^2}{m^2} \Big)^{1-D/2}~.
\label{Dperpintegrals}
}
Since these integrals exist by virtue of dimensional regularization, no pole at  $\ge=0$ arises when expanding them around two dimensions as $D= 2 - 2\ge$. In particular, at the one loop level we often encounter the integral 
\equ{
I_1^0(m^2|D) = \int \frac{\d^D q}{(2\gp)^D \gm^{D-2}}\, 
\frac 1{q^2 + m^2} 
= \frac i{4\gp}\, \gG(\ge)\, \Big( 4\gp \frac{\gm^2}{m^2} \Big)^\ge
= \frac i {4\gp}\, \Big[ 
\frac 1\ge + \ln \frac{\bgm^2}{m^2} 
\Big]~. 
\label{Simple1L}
}

\bibliographystyle{paper}
{\small

\providecommand{\href}[2]{#2}\begingroup\raggedright\endgroup

}

\end{fmffile}
\end{document}